\documentclass[preprint,showpacs,preprintnumbers,amsmath,amssymb]{revtex4}

\usepackage{graphicx}
\usepackage{epsfig}
\usepackage{bm}
\usepackage{amsfonts}
\usepackage{subfigure}
\usepackage{multirow}
\usepackage{color}
\usepackage{float}

\usepackage{environ}         
\usepackage{etoolbox}        
\usepackage{graphicx}        

\newlength{\myl}
\let\origequation=\equation
\let\origendequation=\endequation

\RenewEnviron{equation}{
  \settowidth{\myl}{$\BODY$}                       
  \origequation
  \ifdimcomp{\the\linewidth}{>}{\the\myl}
  {\ensuremath{\BODY}}                             
  {\resizebox{\linewidth}{!}{\ensuremath{\BODY}}}  
  \origendequation
}

\begin{document}

\title{\textbf{Generalized second law of thermodynamics in massive gravity}}

\author{Mohammad Beigmohammadi\footnote{m.beigmohammadi@uok.ac.ir} and Kayoomars Karami\footnote{kkarami@uok.ac.ir}}
\address{\small{Department of Physics, University of Kurdistan, Pasdaran Street, P.O. Box 66177-15175, Sanandaj, Iran}}

\date{\today}

\begin{abstract}
\noindent
Here, we study the generalized second law (GSL) of thermodynamics in the framework of massive gravity.
To do this, we consider a FRW universe filled only with matter and enclosed by the apparent horizon. In addition,
we consider two models including generalized massive gravity (GMG) as well as dRGT massive gravity on de Sitter. For both models, we first study the dynamics of background cosmology and then explore the validity of GSL. We conclude that for the selected values of model parameters the GSL is respected.


\end{abstract}


\maketitle

\newpage
\section{Introduction}\label{sec1}
Recent observational data from the type Ia supernovae (SNeIa) \cite{Riess:1998,Perl:1999},  cosmic microwave background radiation \cite{Sper:2007,Lars:2011} and large-scale structure \cite{Eisen:2005,Perc:2010} provide ample evidence that the current universe is expanding rapidly, and the reason for this acceleration is still unknown.
 A group of cosmologists has tried to include the justification for this acceleration through a strange substance called dark energy into the standard cosmology.
Another group of cosmologists believes in generalized gravity theories. These generalized theories are thought to be gravitational substitutes for dark energy as well as dark matter.

One of the generalized theories of general relativity in the scope of field theory, describing the nonlinear interactions of a spin-2 massive field, is the theory of massive gravity \cite{cai:2012,E.N:2014,Sami:2013,Fierzt:1939,Van:1970,Zak:1970,Va:1972,Dv:2000,
Bou:1972,Rham:2011,Am:2011,Ken:2020,Mic:2020,ruk:2021,Lan:2012,Gon:2013,Gong:2021,
A.E:2011,P.W:2012,T.K:2012,Af:2022,kul:2022,Yus:2022}. Firstly, Fierz and Pauli in 1939 described an action of a free-massive graviton with a second degree mass term \cite{Fierzt:1939}. In addition, in 1970, Van Dam, Veltman and Zakharov discovered that the predictions of linear theory differ from the results of linear general relativity  in the limit of zero mass of the graviton. This difference is known as the vDVZ discontinuity \cite{Van:1970,Zak:1970}. The vDVZ discontinuity in 1972 was rectified by Vainstein, who showed that supplementing the non-interacting Fizir-Pauli theory with a non-linear interaction theory could give rise to smooth this limit \cite{Va:1972,Dv:2000}.

In 1972, Boulware and Deser studied some completely non-linear massive gravity theories and showed that there is a ghost instability in them. Although the linear theory has 5 degrees of freedom, it was found that their investigated nonlinear theories  have 6 degrees of freedom. The extra degree of freedom around non-flat background manifests itself as a scalar field with a kinetic energy that has the wrong sign. This scalar is known as the Boulware-Deser (BD) ghost \cite{Bou:1972}.

Thereafter, in 2011, de Rham, Gabadadze and Tolley (dRGT) founded the dRGT theory, which is free from the sixth extra mode  and the BD ghost \cite{Rham:2011,Am:2011}.
This theory has no closed and flat FRW cosmological solutions. The graviton mass in the open solution, produces an effective cosmological constant.
If one consider the graviton mass of the order of the current
Hubble constant, the effective cosmological constant of massive gravity can justify the current accelerated expansion.

Besides, there are different generalizations of massive gravity such as GMG theory \cite{Ken:2020,Mic:2020,ruk:2021}, and dRGT massive gravity on de Sitter spacetime \cite{Lan:2012,Gong:2021,Gon:2013}. The GMG model is a generalized version of the constant mass dRGT theory in which a slowly varying time dependent mass function is considered for the graviton. This yields the background to remain stable for the allowed region of the parameter space, unlike the standard dRGT model \cite{Ken:2020}. Even with a small change in the model parameters, the strong coupling issue in the dRGT theory becomes more manageable \cite{Mic:2020}. In dRGT massive gravity on de Sitter, the secondary Minkowski metric is replaced by the de Sitter one. This results the model to have a flat cosmological solution in addition to the open solution in ordinary dRGT \cite{Lan:2012,Gong:2021,Gon:2013}. Gumrukcuoglu, Lin, and Mukohyama \cite{A.E:2011} have derived a de Sitter solution with an effective cosmological constant proportional to the mass of the graviton in the dRGT model of massive gravity. The same solution was then obtained for a spatially flat universe in \cite{P.W:2012}. In \cite{T.K:2012}, it was shown that for the specific values of dRGT model parameters, the solution also exists for a universe with arbitrary spatial curvature. Furthermore, in \cite{Af:2022}, inflation has been studied in the dRGT massive gravity framework. In \cite{kul:2022}, the background dynamics and the growth of matter perturbations in the extended quasidilaton setup of massive gravity have been investigated. Additionally, in \cite{Yus:2022}, based on an objective model of massive gravity, the linear growth of matter fluctuations has been studied.

One of another subjects of interest in the cosmology is study of validity of the generalized second law (GSL) of thermodynamics in the accelerating universe. The GSL like the first law of thermodynamics is an accepted law in physics. According to the GSL, the total entropy of the universe including matter and the entropy of horizon should not be reduced over time. The GSL of thermodynamics has been studied in different theories of gravity \cite{karami 1:2011,karami 2:2012,karami 3:2013,karami 4:2013,A.A1:2014,Setare,Basilakos}. The GSL was first formulated by Bekenstein in 1973 for black holes. With the studies done, it was found that the ordinary second law of thermodynamics is violated for black holes. To prevent this violation, the entropy of horizon should be added to the entropy of matter and this is called the generalized second law of thermodynamics. According to that the total entropy of black hole and entropy of the black hole horizon cannot be decreased by increasing the time.

All mentioned in above motivate us to study the GSL of thermodynamics in massive gravity. To this aim, within the framework of massive gravity we consider two models including GMG as well as dRGT on de Sitter and explore the validity of GSL for both of them. The rest of paper is organized as follows: Section \ref{sec2} is devoted to universal thermodynamics in modified gravity. In sections \ref{sec3} and \ref{sec4}, the dynamics of background cosmology and validity of GSL are investigated for the GMG and dRGT on de Sitter models, respectively. At last, the results of our study are summarized in section \ref{sec5}.

\section{Universal thermodynamics}\label{sec2}

For a universe filled with matter and enclosed by the horizon, the GSL is as follows
\cite{Pav:2018,Sa:2014}
\begin{equation}\label{action}
\dot{S}_{\rm tot}=\dot{S_A}+\dot{S}_{m}\geq 0,
\end{equation}
wherein $ S_{A} $ and ${S}_{m} $ denote the horizon and matter entropy, respectively. Also the dot indicates derivative with respect to the cosmic time. To determine $ \dot{S_ A} $ one can use the Clausius relation as
\begin{equation}\label{action}
\delta Q_A=T_{A}dS_A=-d E_A,
\end{equation}
where $ E_A $ is the energy flow across the horizon, and $T_A$ implies the apparent horizon temperature defined as
\begin{equation}\label{T1_eq}
 T_{A} =\frac{1}{2 \pi R_{A}}\left(1-\frac{\dot{R_A}}{2 H R_{A}}\right).
 \end{equation}
Here $H\equiv\dot{a}/a$ is the Hubble parameter, $a$ is the scale factor for expanding universe, and the radius of the apparent horizon $R_{A}$ is as follows
 \begin{equation}\label{R1_eq}
 R_{A}=\frac{1}{\sqrt {H^{2}+\frac{k}{a^2}}},
  \end{equation}
in which $k$ is the spatial curvature. Additionally, the entropy of matter is obtained by the Gibbs equation as follows \cite{Iz:2006}
\begin{equation}\label{F1_eq}
T_{m}dS_{m}= dE_{m}+p_{m}dV,
\end{equation}
in which $E_m=\rho_{m} V$ is the total energy of the matter inside the horizon, $p_{m}$ is the matter pressure,
 $V=\frac{4}{3} \pi R_{A}^3$ is the volume of the universe and $ T_m$ is the matter temperature.

In most generalized gravity theories, the Friedmann equations governing the FRW universe can be written as
\begin{equation}\label{Frid_eq1}
H^2+\frac{k}{a^2}=\frac{8\pi G}{3 }\rho_{t},
\end{equation}
\begin{equation}\label{Frid_eq2}
\dot{H}-\frac{k}{a^2}=-4 \pi G(\rho_{t}+p_{t}),
\end{equation}
wherein  $\rho_{t}=\rho_m+\rho_{DE} $ and $ p_{t}=p_m+p_{DE} $ indicate the total energy density and pressure. Here $(\rho_m, p_m)$ are related to the distribution of matter and ($\rho_{DE}$,$ p_{DE}$) can be considered as effective density and pressure of dark energy.

Applying the Clausius relation $\delta Q_A=T_{A}dS_A$, the apparent horizon entropy is computed as follows \cite{Mitra:2014}
\begin{equation}\label{s1_eq}
 S_{A} =\frac{A}{4G}-8 \pi^2 \int H R_{A}^4 (\rho_{DE}+p_{DE}) dt,
\end{equation}
in which $A=4\pi R_{A}^2 $ denotes the horizon area. Equation (\ref{s1_eq}) shows that the apparent horizon entropy is the standard Bekenstein entropy with a modified term (in integral form).

From Eqs. (\ref{T1_eq}), (\ref{F1_eq}) and using
$\dot{\rho}_m+3H(\rho_m+p_m)=0$, the entropy of matter inside the universe can be computed as follows
\begin{equation}\label{s2_eq}
T_{A}\dot{S_{m}}=4\pi R_{A}^2(\rho_{m}+p_{m})\Big[\dot{R_{A}}-H R_{A}\Big],
\end{equation}
where we have assumed the horizon and fluid temperature are equivalent to each other $T_{m}=T_{A}$.

From Eq. (\ref{s1_eq}) one can calculate the entropy of the apparent horizon as
\begin{equation}\label{a2_eq}
T_{A}\dot{S_{A}}=\frac {1}{2 \pi R_{A}}\left(1-\frac{\dot{R}_A}{2 H R_A}\right) \left[\frac{2\pi R_A \dot{R}_A}{G} -8\pi^2 H R_A^4(\rho_{DE} + p_{DE})\right].
\end{equation}

Using Eqs. (\ref{R1_eq}) and (\ref{Frid_eq2}), the time derivative of the apparent horizon radius is obtained as follows
\begin{eqnarray}\label{r1_eq}
\dot{R_A}&=&-HR_{A}^3\left(\dot{H}-\frac{k}{a^2}\right),\nonumber\\
&=&4\pi G H R_{A}^3 \big[\rho_m+p_m+(\rho_{DE}+p_{DE})\big].
\end{eqnarray}
Replacing Eq. (\ref{r1_eq}) into (\ref{s2_eq}) and (\ref{a2_eq}) one can get
\begin{equation}\label{s11_eq}
T_A\dot{S_m}=16\pi^2 G H R_A^5 (\rho_m+p_m) \big[\rho_m+p_m+(\rho_{DE}+p_{DE})\big]-4\pi H R_A^3\left(\rho_m+p_m\right),
\end{equation}

\begin{equation}\label{s12_eq}
T_A\dot{S_{A}}=4 \pi H R_{A}^3(\rho_m+p_m)-8\pi^2 G H R_{A}^5(\rho_m+p_m) \big[\rho_m+p_m+ (\rho_{DE}+p_{DE})\big].
\end{equation}
Adding Eqs. (\ref{s11_eq}) and (\ref{s12_eq}), the GSL of thermodynamics in modified gravity can be obtained as
\begin{equation}\label{st_eq}
T_A\dot{S}_{\rm tot}=T_A\dot{S_{A}}+T_A\dot{S_m}=8\pi^2 G H R_{A}^5 (\rho_m+p_m)\big [\rho_m+p_m+(\rho_{DE}+p_{DE})\big].
\end{equation}
Note that in the Einstein gravity, where $S_A=\frac{A}{4\pi G}$ from Eq. (\ref{s1_eq}) we have $\rho_{DE}+p_{DE}=0$ and consequently, Eq. (\ref{st_eq}) reduces to
\begin{equation}
T_{A} \dot{S}_{\rm tot}= 8\pi^2 G H R_{A}^5 (\rho_m+p_m)^2\geq 0,
\end{equation}
which shows that the GSL is always respected in the Einstein gravity \cite{Akbar:2008}.

In the next sections, within the framework of massive gravity, we consider the two models including the GMG and dRGT on de Sitter and examine the validity of the GSL for both models.
\section{Generalized Massive Gravity}\label{sec3}
The action of GMG model including the Einstein-Hilbert action and the generalized dRGT takes the form \cite{Fa:2014}
\begin{equation}\label{Lagrangian1}
S= M_{P}^2\int d^4x\sqrt{-g}\left[\frac{R}{2}+m^2\sum^{4}_{n=0}\alpha_{n}(\phi^a\phi_{a})U_{n}[\mathcal{K}]\right]+S_m,
\end{equation}
in which $M_{P}=1/\sqrt{8\pi G}$ is the reduced Planck mass, $g$ is determinant of the metric tensor, $R$ is the Ricci scalar, $m$ is the graviton mass and $S_m$ is the action of matter. Also the free parameters
$\alpha_n(\phi^a\phi_a)$ are functions of the Lorentz invariant term $\eta_{ab}\phi^{a}\phi^{b}$ in which $\phi^a$ is the St\"{u}ckelberg field.

Also the functions $ U_n$ in the action (\ref{Lagrangian1}) denote the dRGT potential terms which are defined as
\begin{align}\label{un}
&U_0 (\mathcal{K})=1,\nonumber\\
&U_1 (\mathcal{K})=[\mathcal{K}],\nonumber\\
&U_2 (\mathcal{K})=\frac{1}{2!}([\mathcal{K}]^2-[\mathcal{K}^2]),\nonumber\\
&U_3 (\mathcal{K})=\frac{1}{3!}([\mathcal{K}]^3-3[\mathcal{K}][\mathcal{K}^2]+2[\mathcal{K}^3]),\nonumber\\
&U_4 (\mathcal{K})=\frac{1}{4!}([\mathcal{K}]^4-6[\mathcal{K}]^2[\mathcal{K}^2]+8[\mathcal{K}][\mathcal{K}^3]+3[\mathcal{K}^2]^2-6[\mathcal{K}^4]),
\end{align}
where the square brackets denote the trace operation on the tensor $\mathcal{K}$ defined as
\begin{equation}
\mathcal{K}^{\mu}_{\nu}=\delta^{\mu}_{\nu} -\big(\sqrt{g^{-1} f}~\big)^\mu_\nu.
\end{equation}
Also $f_{\mu\nu}$ is the fiducial metric which is defined in terms of four St\"{u}ckelberg fields $\phi^a$ as $f_{\mu\nu}\equiv\eta_{ab}\partial_{\mu}\phi^a\partial_{\nu}\phi^b$.

Here, we consider an open FRW background as
\begin{equation}\label{gmunu}
g_{\mu\nu}dx^{\mu}dx^{\nu}=-dt^{2}+a(t)^2\Omega_{ij}dx^{i}dx^{j},
\end{equation}
which $\Omega_{ij}$ is the spatial metric given by
\begin{equation}\label{Lagrangian}
\Omega_{ij}dx^{i}dx^{j}=dx^{2}+dy^{2}+dz^{2}-\frac{k(xdx+ydy+zdz)^2}{1+k(x^{2}+y^{2}+z^{2})},
\end{equation}
and $k= |K|=-K$ is the negative absolute value of the constant spatial curvature.

For the above background, the configuration of the St\"{u}ckelberg field corresponding to the conditions of homogeneity and isotropy takes the form \cite{Mic:2020,Lin:2011}
\begin{align}\label{stuk_eq}
&\phi^{0}=f(t)\sqrt{1+k(x^2+y^2+z^2)},\nonumber\\
&\phi^{1}=f(t)\sqrt{k}~x,\nonumber\\
&\phi^{2}=f(t)\sqrt{k}~y,\nonumber\\
&\phi^{3}=f(t)\sqrt{k}~z,
\end{align}
where the field configuration (\ref{stuk_eq}) satisfy the relation $\phi^a\phi_a=-f(t)^2$.

Using Eq. (\ref{stuk_eq}),  the fiducial metric $f_{\mu\nu}=\eta_{ab}\partial_{\mu}\phi^a\partial_{\nu}\phi^b$ corresponding to the Minkowski spacetime $\eta_{ab}$ in an open universe reads
\begin{equation}\label{fmunu}
f_{\mu\nu} dx^{\mu}dx^{\nu}=-\dot{f}(t)^2 dt^2+k f(t)^2\Omega_{ij} dx^{i}dx^{j},
\end{equation}
where $f(t)$ is the St\"{u}ckelberg field function.

For the matter part, we consider the energy-momentum tensor of the perfect fluid as
\begin{equation}
T_{\mu\nu}=\rho_m u_{\mu}u_{\nu}+p_m(g_{\mu\nu}+u_{\mu}u_{\nu}),
\end{equation}
where $u_{\mu}$ is the four velocity field. In what follows, we consider the pressureless matter ($p_m=0$) throughout the paper.
Therefore, the continuity relation $\dot{\rho}_m+3H\rho_m=0$ governing the pressureless matter yields $\rho_m=\rho_{m_0}a^{-3}$.

Varying the action (\ref{Lagrangian1}) with respect to the physical metric $g_{\mu\nu}$, Eq. (\ref{gmunu}), one can get the modified Friedmann equations
\begin{equation}\label{frid1_eq}
3\left(H^2-\frac {k}{a^2}\right)=m^2L+\frac{\rho_{m}}{M_P^2},
\end{equation}
\begin{equation}\label{frid2_eq}
2\left(\dot{H}+\frac{k}{a^2}\right)=m^2J(r-1)\xi-\frac{\rho_m}{M_P^2},
\end{equation}
 where
\begin{align}\label{LJ}
& L\equiv-\alpha_{0}+(3\xi-4)\alpha_{1}-3(\xi-1)(\xi-2)\alpha_{2}+(\xi-1)^2(\xi-4)\alpha_3+(\xi-1)^3\alpha_{4},\nonumber\\
&J\equiv\alpha_{1}+(3-2\xi)\alpha_{2}+(\xi-1)(\xi-3)\alpha_{3}+(\xi-1)^2\alpha_{4},
 \end{align}
and
\begin{equation}\label{all_eq}
\qquad\xi\equiv\frac{\sqrt {k} f}{a}, \qquad r\equiv\frac{a\dot f}{\sqrt{k}f}.
\end{equation}
Taking variation of the action (\ref{Lagrangian1}) in terms of the fiducial metric $f_{\mu\nu}$, Eq. (\ref{fmunu}), one can obtain the background St\"{u}ckelberg equation
\begin{equation}\label{stu2_eq}
3HJ\left(r-1\right)\xi-\dot{L}=0.
\end{equation}
Following \cite{Ken:2020,Mic:2020} for the minimal model of GMG, the functions $\alpha$ take the forms
\begin{align}\label{alpha_eq}
& \alpha_{0}(\phi^a\phi_{a}) =\alpha_1(\phi^a\phi_{a})=0,\nonumber\\
& \alpha_{2}(\phi^a\phi_{a})=1+m^2\alpha'_2\phi^a\phi_{a},\nonumber\\
&\alpha_3(\phi^a\phi_{a})=\alpha_3,\nonumber\\
&\alpha_4(\phi^a\phi_{a})=\alpha_4,
\end{align}
where $\alpha'_2$, $\alpha_3$ and $\alpha_4$ are the free model parameters.

\begin{figure}[H]
\begin{minipage}[b]{1\textwidth}
\subfigure[\label{fig-phi} ]{ \includegraphics[width=.49\textwidth]%
{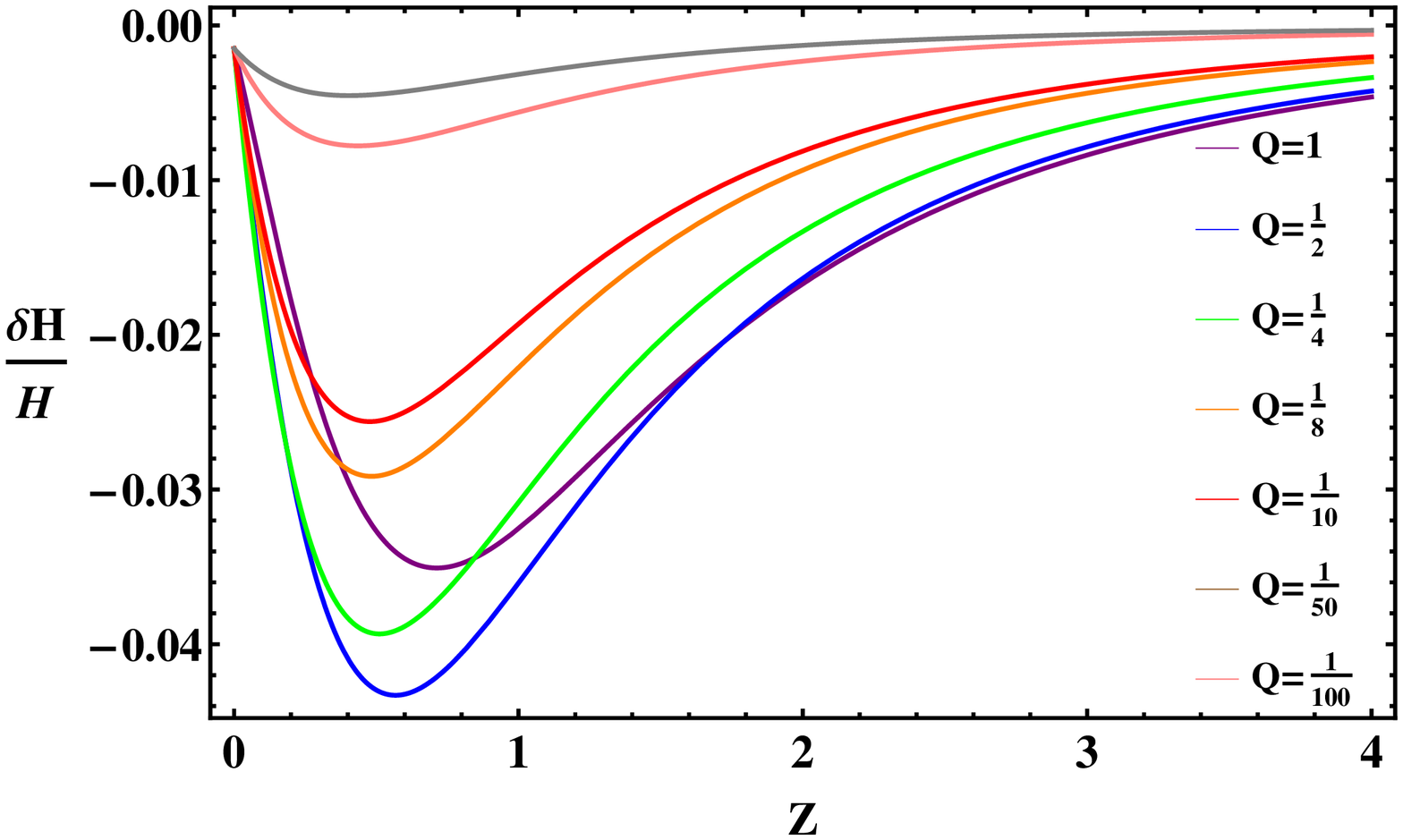}}\hspace{.1cm}
\subfigure[\label{fig-phi} ]{ \includegraphics[width=.49\textwidth]%
{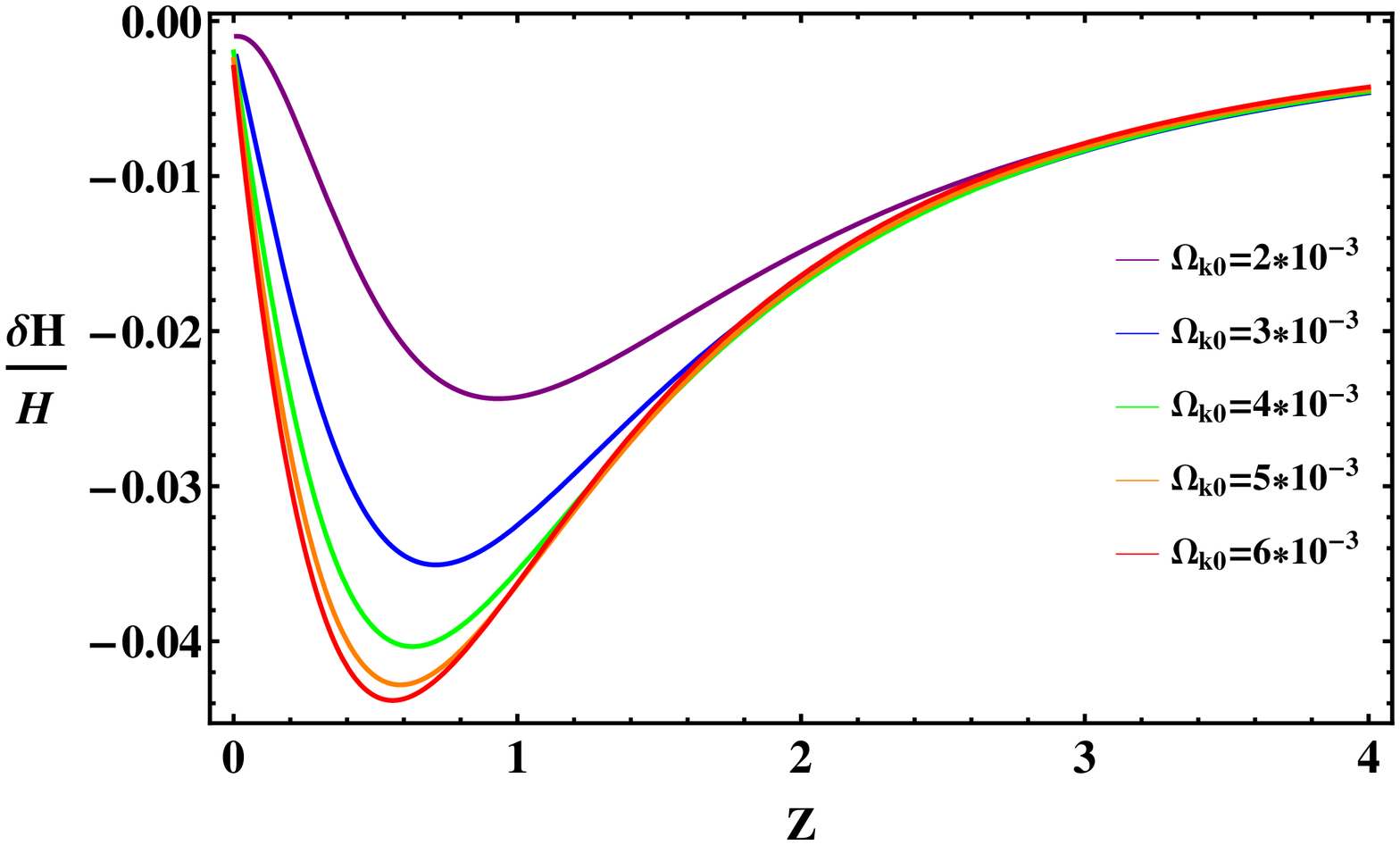}}\\
\subfigure[\label{fig-srp}]{ \includegraphics[width=.48\textwidth]%
{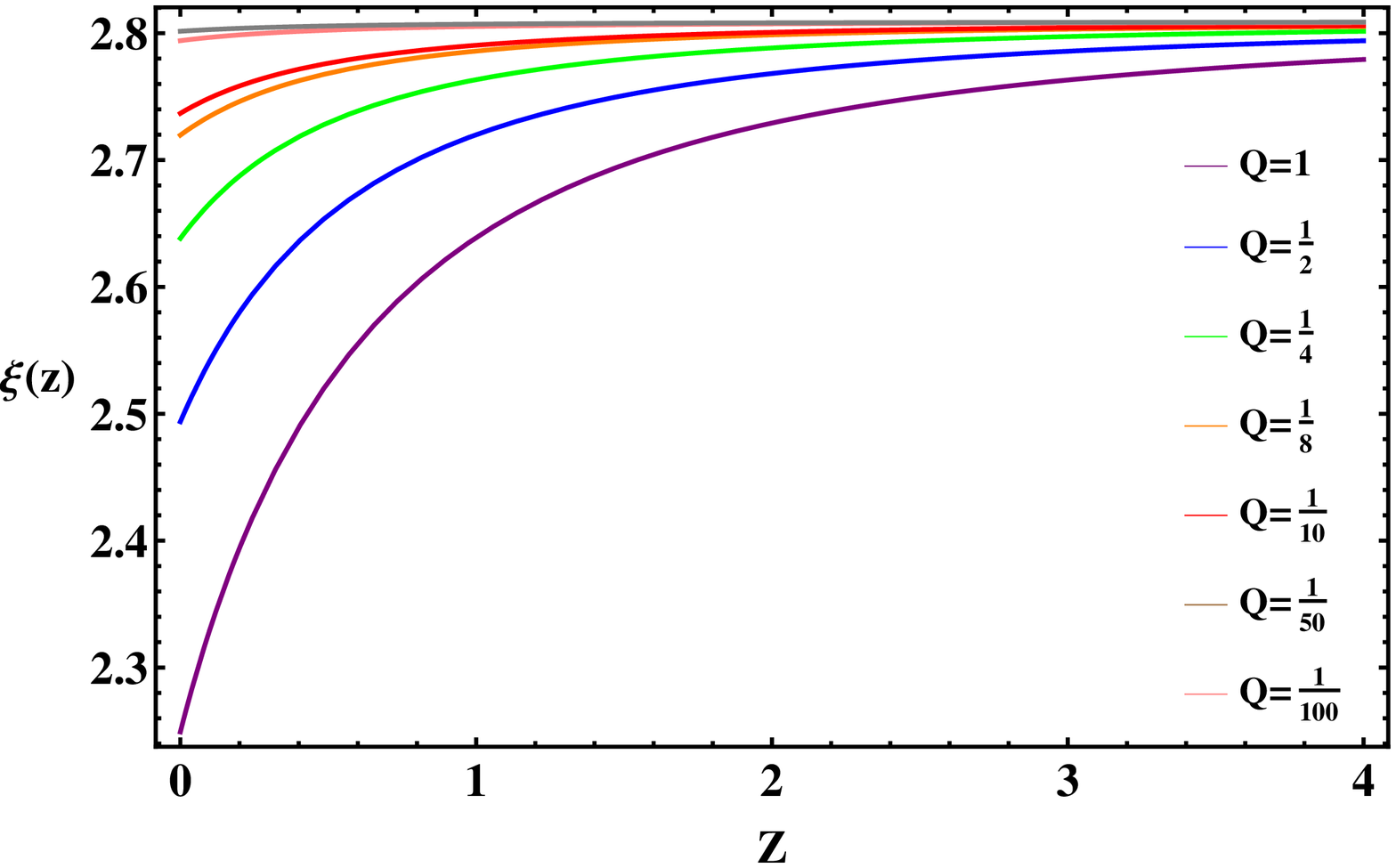}}\hspace{.1cm}
\subfigure[\label{fig-srp}]{ \includegraphics[width=.48\textwidth]%
{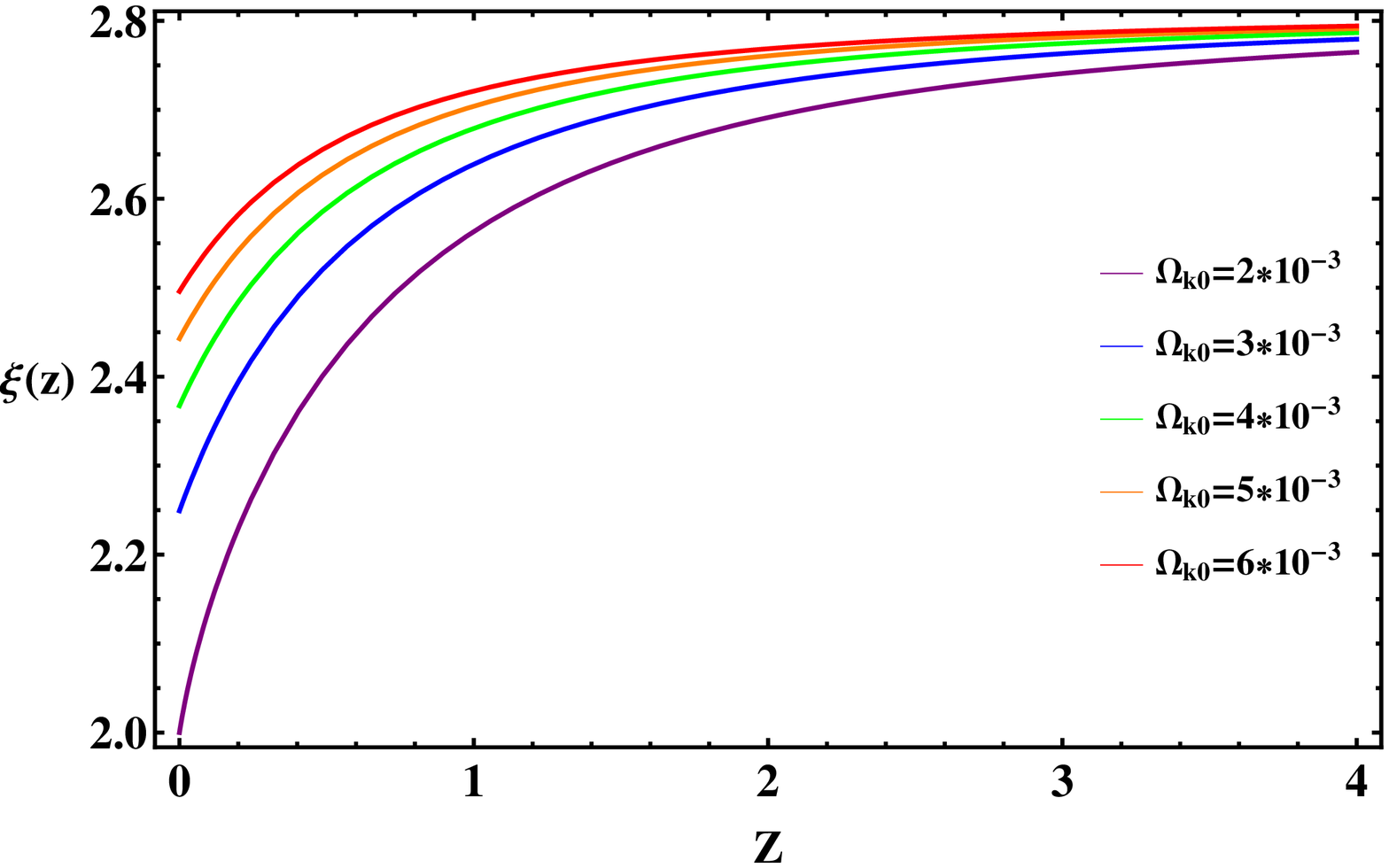}}
\end{minipage}
\caption{Evolutions of the fractional deviation of the Hubble parameter $\frac{\delta H}{H}=\frac{H-H_{\Lambda\rm CDM}}{H_{\Lambda \rm CDM}}$ in GMG model from $\Lambda$CDM as well as the quantity $\xi(z)$ for different values of $Q$ (with fixed $\Omega_{k_0}=3\times 10^{-3}$) and $\Omega_{k_0}$ (with fixed $Q=1$). Auxiliary parameters are $\Omega_{m_0}=0.3$, $\alpha_{3}=0$ and $\alpha_{4}=0.8$ \cite{Mic:2020}.  Here $H_{\Lambda \rm CDM}=H_{0}\sqrt{\Omega_{m_0}(1+z)^3+\Omega_{k_0}(1+z)^2+\Omega_{\Lambda_0}}$.
  }\label{linear1}
\end{figure}

Using the following dimensionless parameters
\begin{align}\label{dimenlessparam}
& m\equiv H_{0}\mu,\nonumber\\
& H\equiv H_{0}h,\nonumber\\
& k\equiv a_{0}^2H_{0}^2\Omega_{k_0},\nonumber\\
& \alpha'_{2}\equiv \frac{10^{-4}Q}{\mu^2},\nonumber\\
& \rho_m\equiv \frac {3a_0^3 H_0^2 M_P^2\Omega_{m_0}}{a^3},
\end{align}
and with the help of Eq. (\ref{all_eq}), the first Friedmann equation (\ref{frid1_eq}) and the St\"{u}ckelberg equation (\ref{stu2_eq}) can be recasted in the following form
\begin{align}\label{fre1_eq}
& 3\left(h^2-\frac{a_0^2\Omega_ {k_0}}{a^2}\right)-\mu^2\Bigg[\frac{3(\xi-2)(\xi-1)(10^{-4}Q a^2 \xi^2-a_0^2\Omega_ {k_0})}{a_0^2\Omega_ {k_0}}\nonumber\\
&+\alpha_3(\xi-1)^2 (\xi-4)+\alpha_4(\xi-1)^3\Bigg]=\frac{3a_0^3\Omega_ {m_0}}{a^3},
\end{align}

\begin{align}\label{fre10_eq}
&3\left(h-\frac{\sqrt{\Omega_{k_0}}a_0}{a}\right)\Bigg[(3-2\xi)\left(1-\frac{10^{-4}Q a^2 \xi^2}{a_0^2 \Omega_ {k_0}}\right)\nonumber\\
&+\alpha_3(\xi-3)(\xi-1)+\alpha_4(\xi-1)^2\Bigg]\nonumber\\
&=\frac{6 (10^{-4}Q) a(\xi-2)(\xi-1)\xi}{a_0\sqrt{\Omega_ {k_0}}}.
\end{align}
The evolution of $ h$ and $ \xi $ can be determined by solving the above equations. Using the St\"{u}ckelberg equation (\ref {fre10_eq}), we obtain a solution for $h$. Replacing this into the first Friedmann equation (\ref{fre1_eq}) yields a tenth order polynomial equation for $\xi$ and to select the physical solution with positive real values, we compare the roots of the equation to the value of $\xi_{\rm dRGT}=2.80902$ at early times. The results are shown in Fig. \ref{linear1}.
Figure \ref{linear1} shows that: (i) with decreasing $Q$, $\delta H/H$ goes to zero and the GMG model behaves like $\Lambda$CDM one. (ii) For a given $z$ when $\Omega_{k_0}$ increases, both $|\delta H/H|$ and $\xi(z)$ increase. (iii) For a given $z$, when $Q$ decreases, $\xi(z)$ increases and goes to $\xi_{\rm dRGT}=2.80902$. This shows that the GMG model for $Q\rightarrow 0$ recovers the result of standard dRGT massive gravity.

In the next, comparing the Friedmann Eqs. (\ref{frid1_eq}) and (\ref{frid2_eq}), respectively, with Eqs. (\ref{Frid_eq1}) and (\ref{Frid_eq2}) one can obtain
\begin{equation}\label{rhode}
\rho_{DE}=M^2_Pm^2L,
\end{equation}
\begin{equation}\label{L2}
p_{DE}+\rho_{DE}=-M_{P}^2 m^2 J (r-1)\xi.
\end{equation}
Using Eqs. (\ref{rhode}) and (\ref{L2}), the effective equation of state parameter of dark energy can be written as follows
\begin{equation}\label{om1_eq}
\omega_{DE}\equiv\frac{p_{DE}}{\rho_{DE}}=-1-\frac{\ J (r-1)\xi}{L}.
\end{equation}
From Eq. (\ref{all_eq}), the quantity $r$ can be expressed in terms of $H$ and $\xi$ as
\begin{equation}
r=\frac{a}{\sqrt{k}}\Bigg(H+\frac{\dot{\xi}}{\xi}\Bigg).
\end{equation}
With the help of Eqs. (\ref{alpha_eq}) and (\ref{dimenlessparam}) and using  $\phi^a\phi_a=-f(t)^2$, the dimensionless parameter $\alpha_2$ takes the form
 \begin{equation}\label{alpha2}
\alpha_2(t)=1-\frac{Q}{10^4} H_0^2f(t)^2.
\end{equation}
Note that the parameter $\alpha_2$ shows the departure of GMG model from the dRGT one and for the case of $Q\rightarrow 0$ we have $\alpha_2=1$ and consequently the GMG model reduces to the standard dRGT massive gravity as shown in Fig. \ref{linear1}.

 From Eq. (\ref{rhode}) and using Eqs. (\ref{LJ}), (\ref{all_eq}), (\ref{dimenlessparam}), (\ref{alpha2}), the effective density parameter of dark energy arising from the mass term in the GMG model can be obtained as follows
\begin{eqnarray}\label{omga1_eq}
\Omega_{DE}&\equiv&\frac{\rho_{DE}}{3M_P^2H^2}=\frac{\mu^2}{3h^2}L,\nonumber\\
&=&\frac{\mu^2}{3h^2}(\xi-1)\left[\alpha_3(\xi-1)(\xi-4)+\alpha_4(\xi-1)^2+3(\xi-2)\left(\frac{Q\xi^2a^2}{10^4\Omega_{k_0}a_0^2}-1\right)\right],
\end{eqnarray}
where we have used $\alpha_0=\alpha_1=0$ from Eq. (\ref{alpha_eq}). Here, the value of $\mu$ parameter can be determined by applying the constraint $\Omega_{k_0}+\Omega_{m_0}+\Omega_{DE_0}=1$ at the present time.

\begin{figure}[H]
\begin{minipage}[b]{1\textwidth}
\subfigure[\label{fig-phi} ]{ \includegraphics[width=.48\textwidth]%
{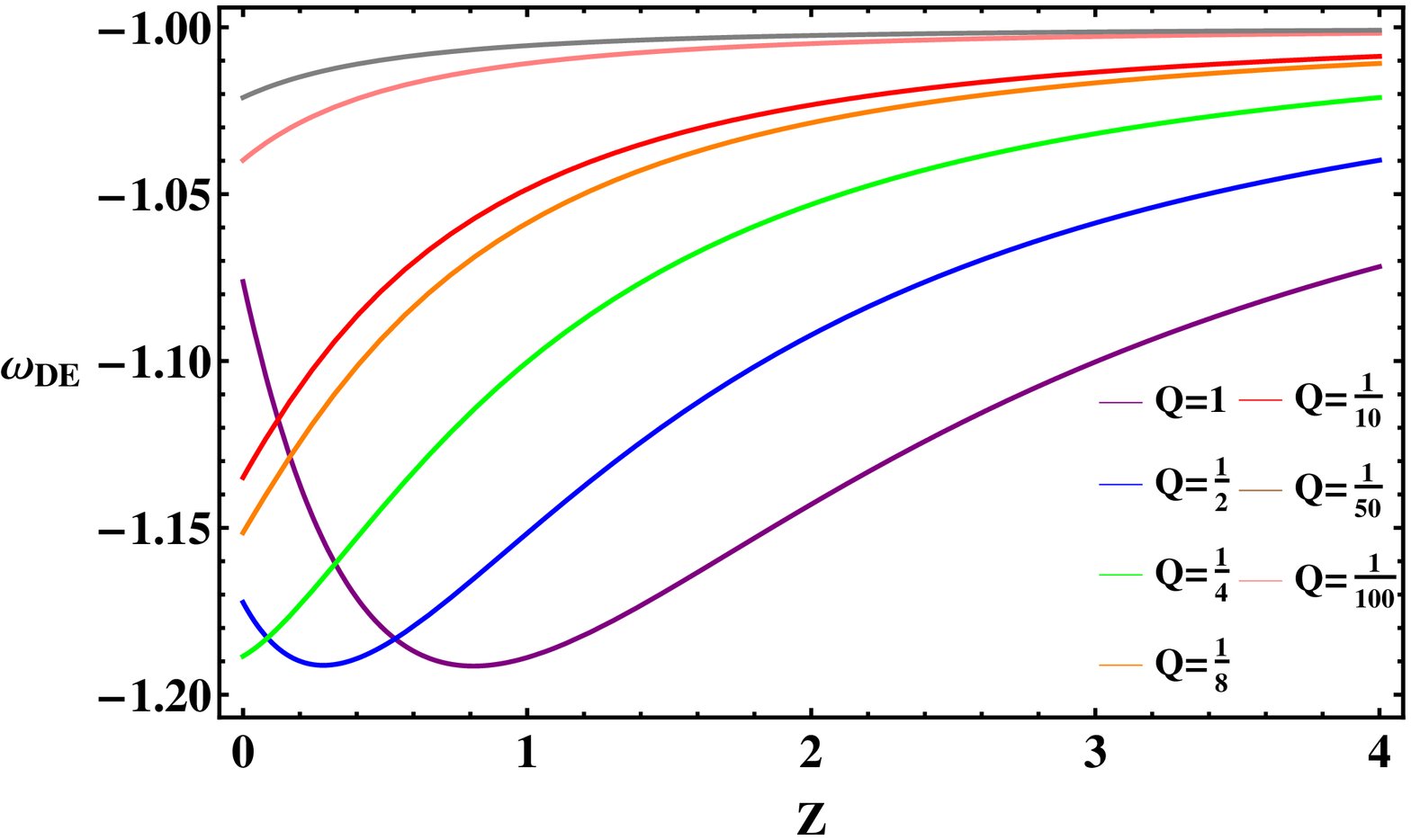}}\hspace{.1cm}
\subfigure[\label{fig-srp}]{ \includegraphics[width=.48\textwidth]%
{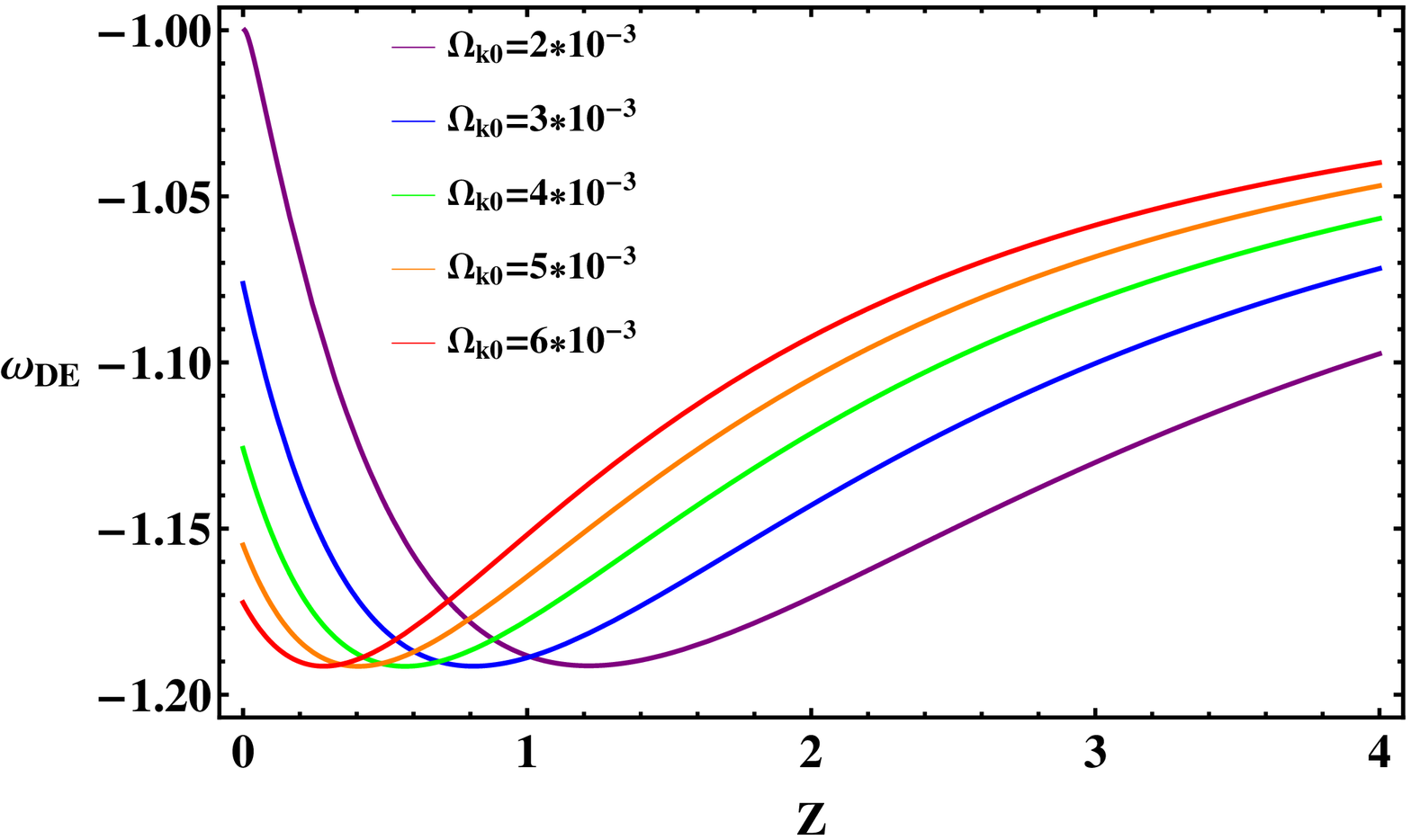}}\\
\subfigure[\label{fig-srp}]{ \includegraphics[width=.49\textwidth]%
{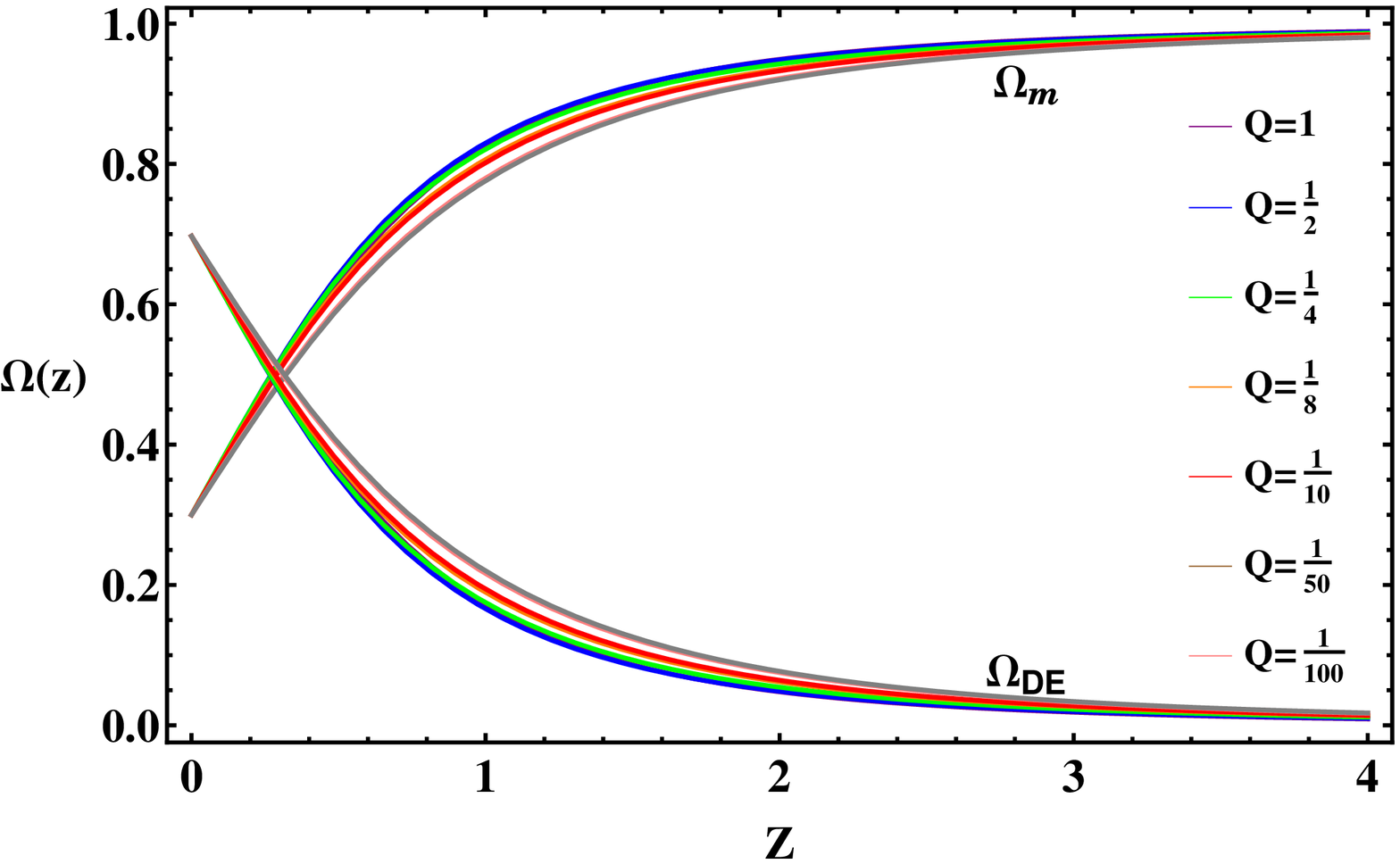}}\hspace{.1cm}
\subfigure[\label{fig-srp}]{ \includegraphics[width=.49\textwidth]%
{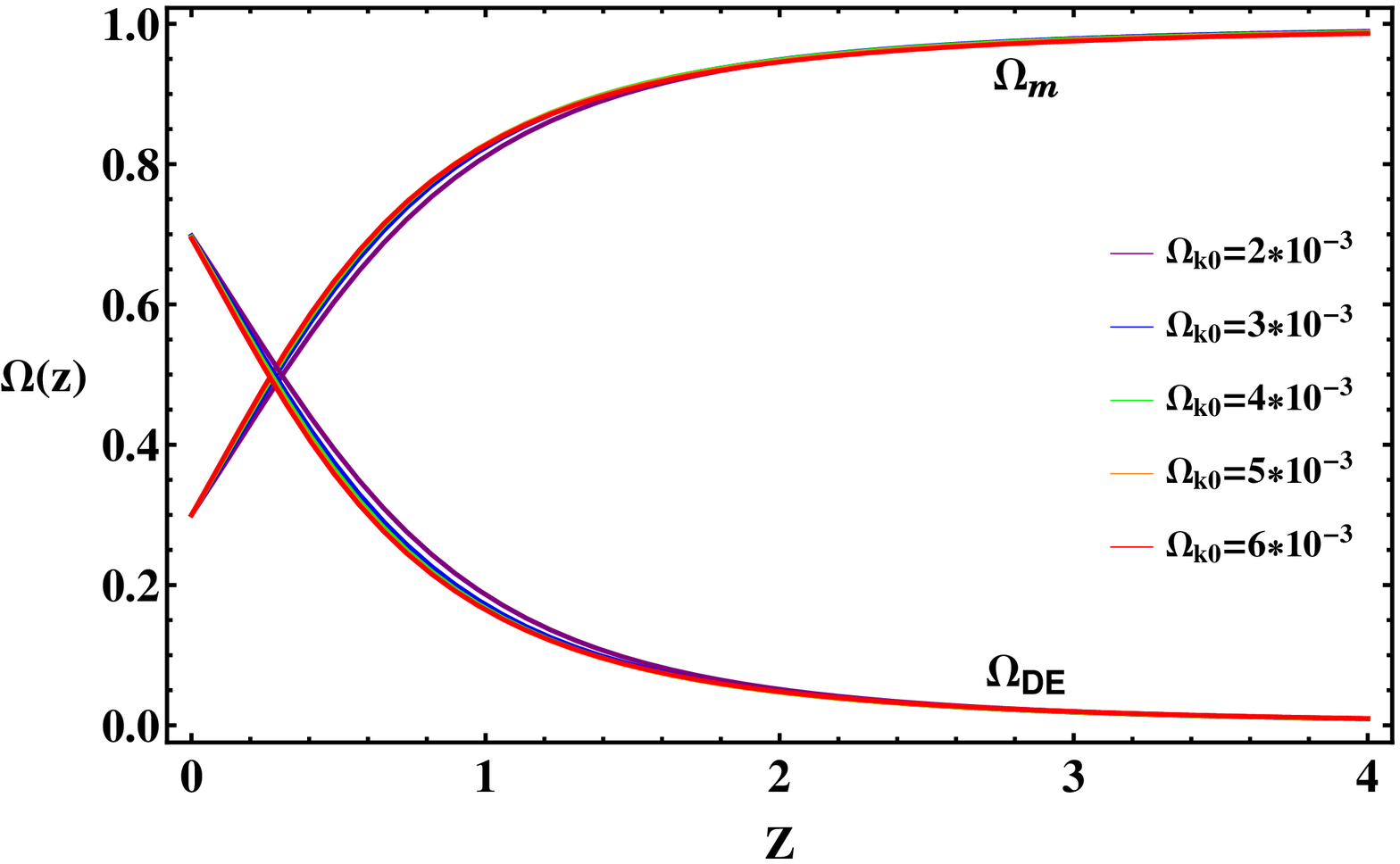}}\\
\subfigure[\label{fig-srp}]{ \includegraphics[width=.49\textwidth]%
{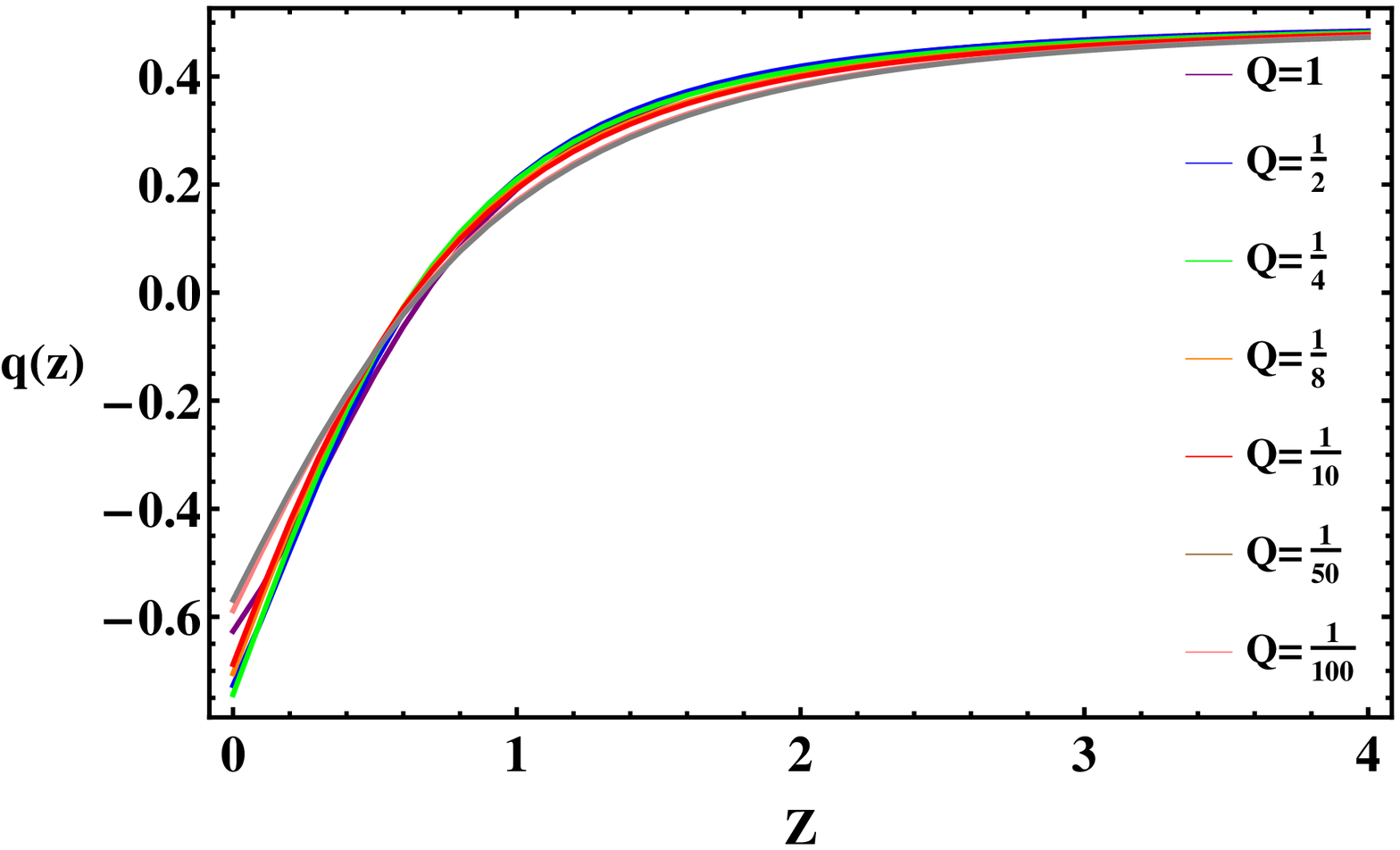}}\hspace{.1cm}
\subfigure[\label{fig-srp}]{ \includegraphics[width=.49\textwidth]%
{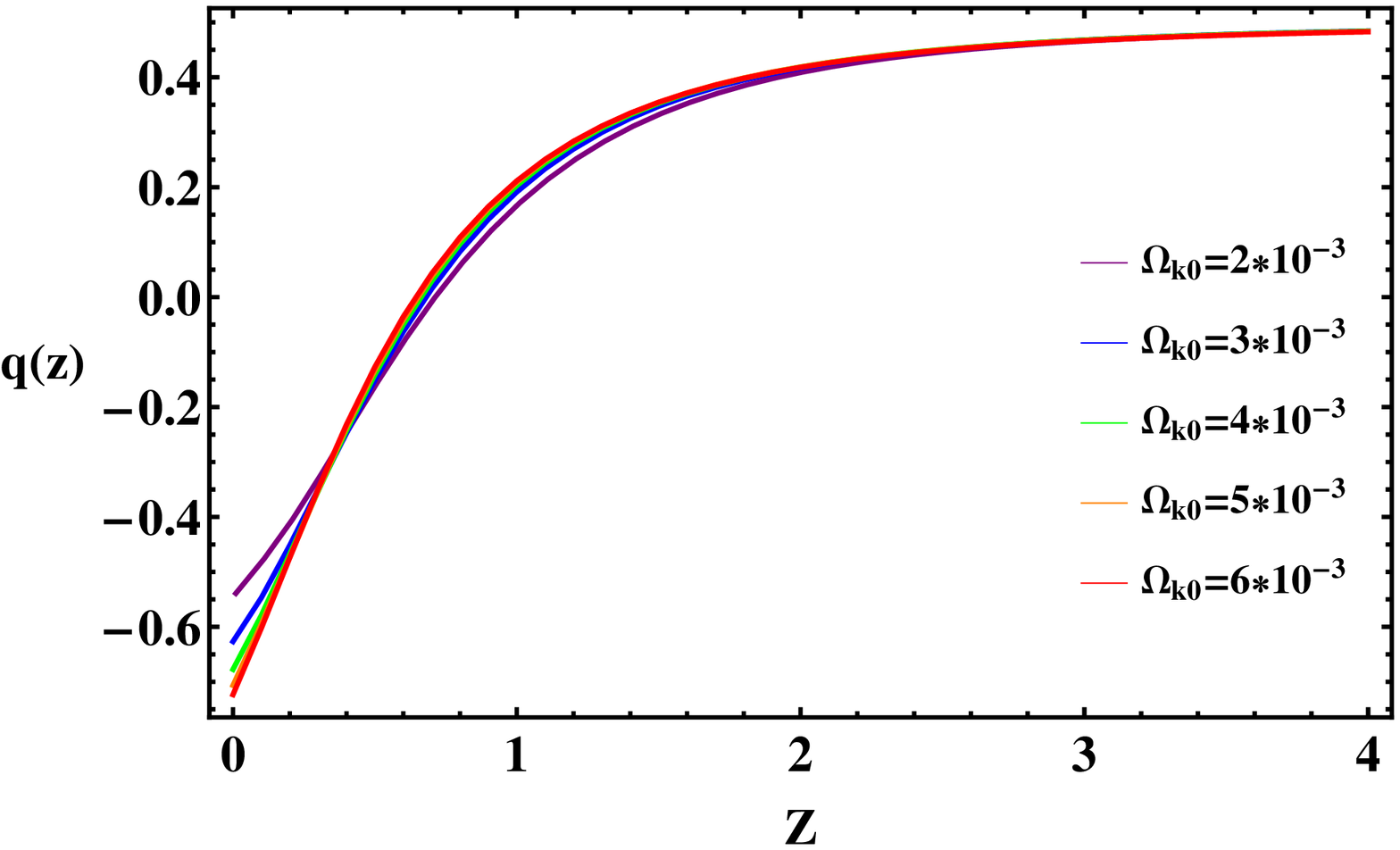}}
\end{minipage}
\caption{Variations of the equation of state parameter of dark energy $\omega_{DE}$, the density parameters  ($\Omega_{m}$,$\Omega_{DE}$) and the deceleration parameter $q(z)$ versus the redshift $z$ for GMG model with different values of $Q$ (with fixed $\Omega_{k_0}=3\times 10^{-3}$) and $\Omega_{k_0}$ (with fixed $Q=1$). Auxiliary parameters are $\Omega_{m_0}=0.3$, $\alpha_{3}=0$ and $\alpha_{4}=0.8$ \cite{Mic:2020}.
  }\label{linear}
\end{figure}

With the help of results of $h$ and $\xi$ obtained from numerical solving of Eqs. (\ref{fre1_eq}) and (\ref{fre10_eq}), one can get the evolutions of $\omega_{DE}$, $\Omega_{DE}$, $\Omega_{m}=\Omega_{m_0}a^{-3}/h^2$ and the deceleration parameter  $q(z)=-1-\dot{H}/H^2$ versus the redshift $z=\frac{a_0}{a}-1$ for different set of model parameters. The results are illustrated in Fig. \ref{linear} which shows that (i) the equation of state parameter of GMG model behaves like phantom dark energy, i.e. $\omega_{DE}<-1$, for different $Q$ and $\Omega_{k_0}$. (ii) For the case of $Q\rightarrow 0$ we have $\omega_{DE}\rightarrow -1$ and the model behaves like the $\Lambda$CDM one. (iii) The density parameter of matter $\Omega_{m}$ and dark energy $\Omega_{DE}$ start, respectively, from 1 and zero and then tend to their values at the present time. (iv) The deceleration parameter $q(z)$ begins from matter dominated universe, i.e. $q=0.5$, and then shows a transition from decelerating ($q>0$) to accelerating ($q<0$) universe in the recent past. (v) Both $\Omega(z)$ and $q(z)$ for $Q\rightarrow 0$ behave like the $\Lambda$CDM model.

\subsection{Generalized second law of thermodynamic in GMG model}\label{subsec1}
\begin{figure}[H]
\begin{minipage}[b]{1\textwidth}
\subfigure[\label{fig-phi} ]{ \includegraphics[width=.48\textwidth]%
{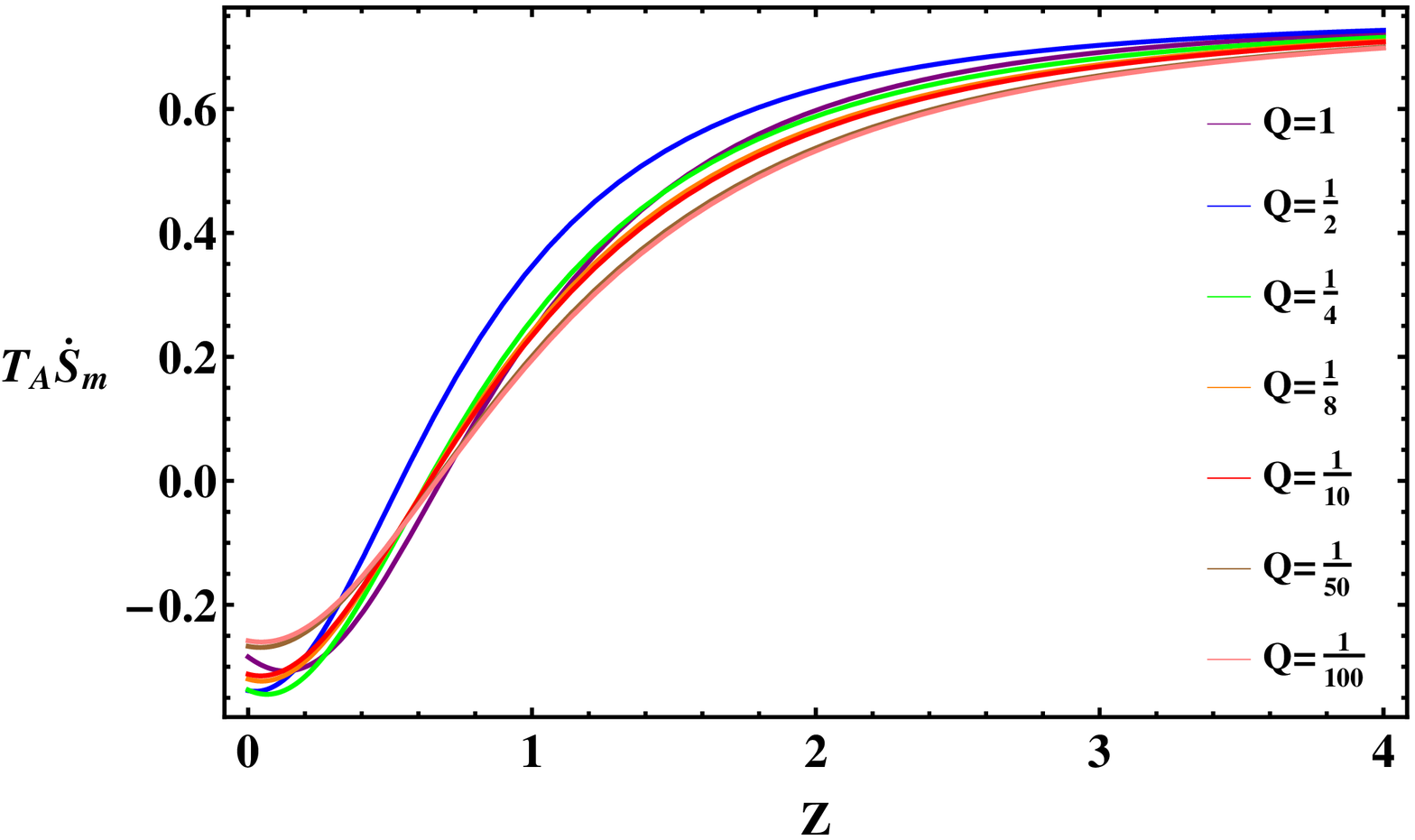}}\hspace{.1cm}
\subfigure[\label{fig-srp}]{ \includegraphics[width=.49\textwidth]%
{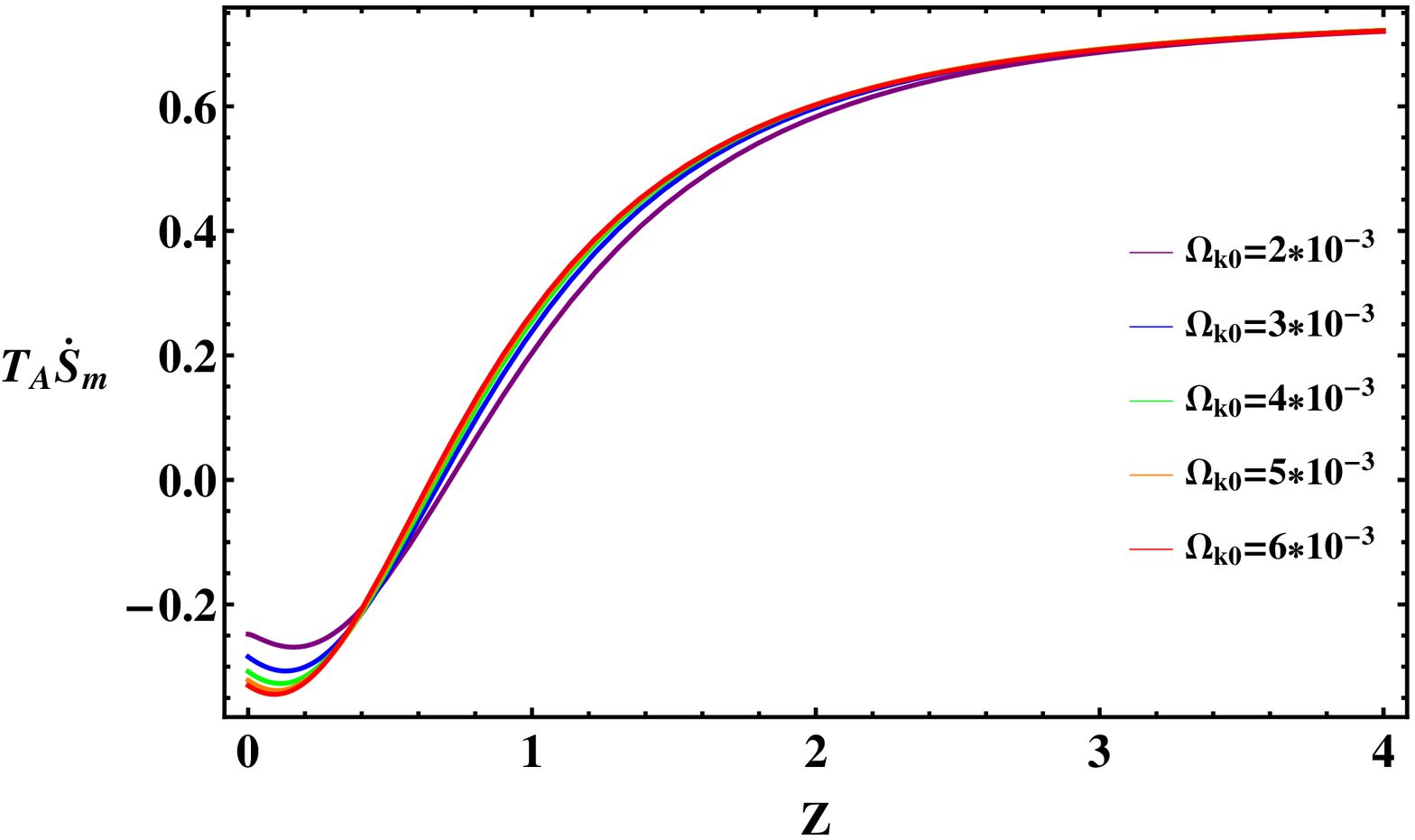}}\\
\subfigure[\label{fig-srp}]{ \includegraphics[width=.48\textwidth]%
{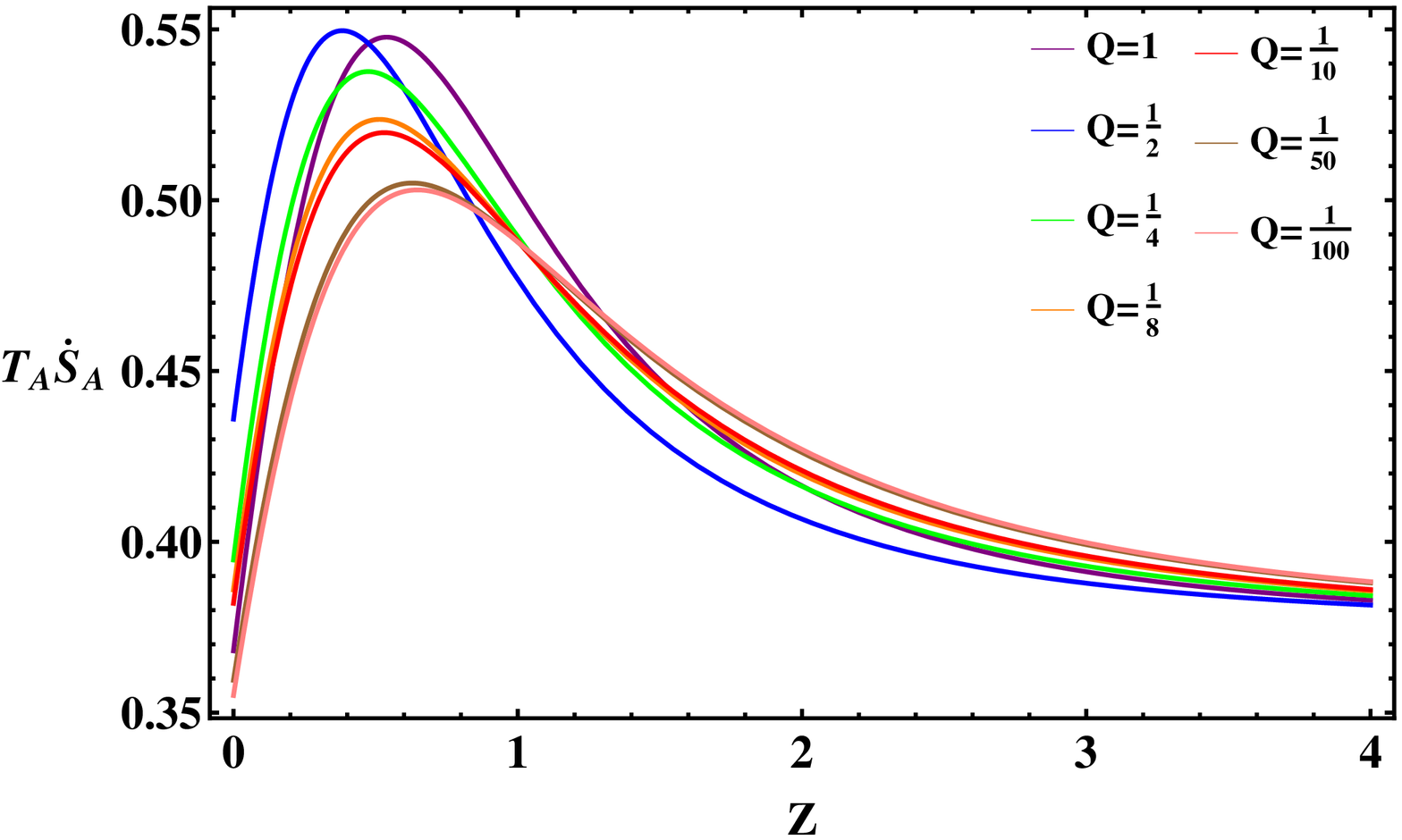}}\hspace{.1cm}
\subfigure[\label{fig-srp}]{ \includegraphics[width=.49\textwidth]%
{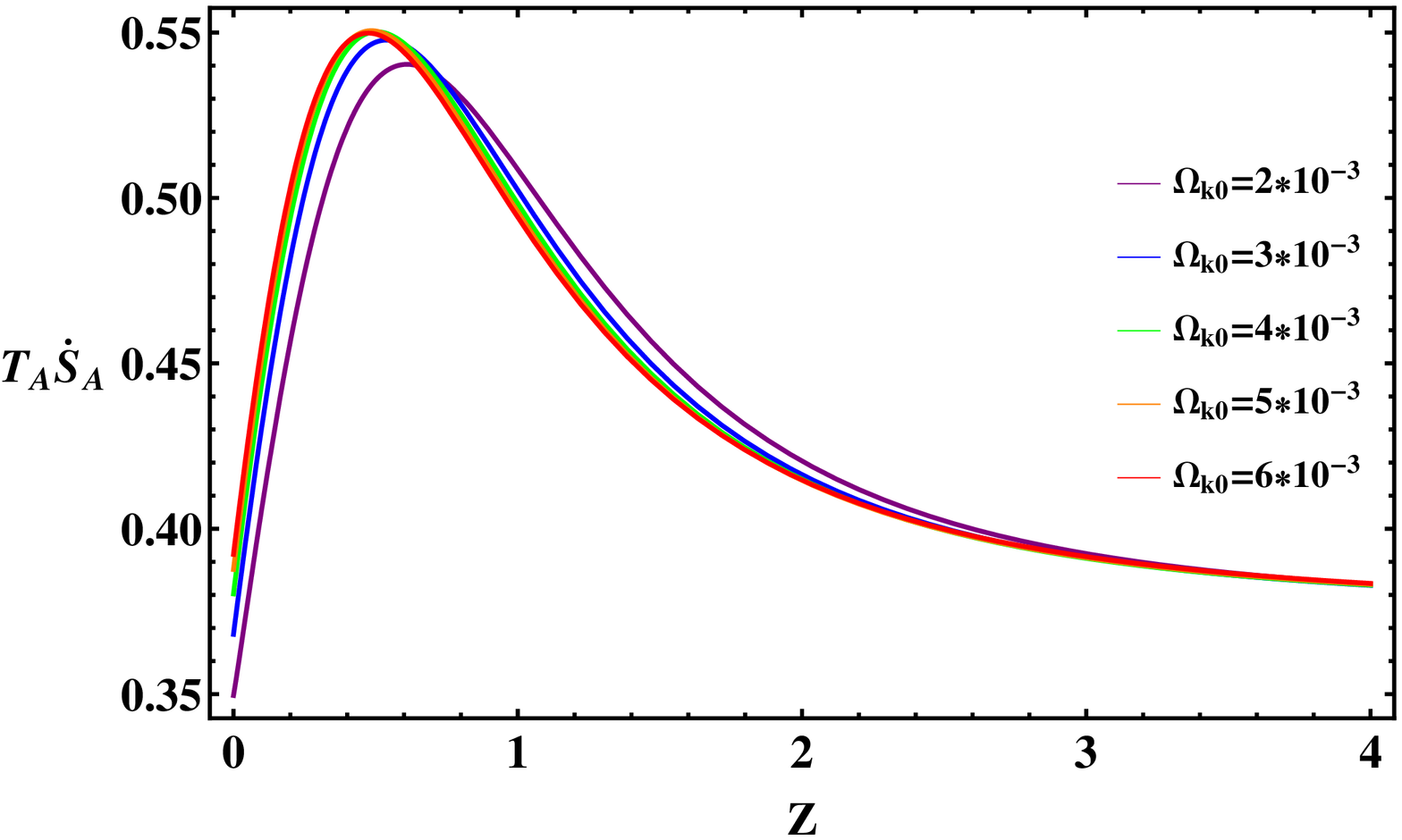}}\\
\subfigure[\label{fig-phi} ]{ \includegraphics[width=.48\textwidth]%
{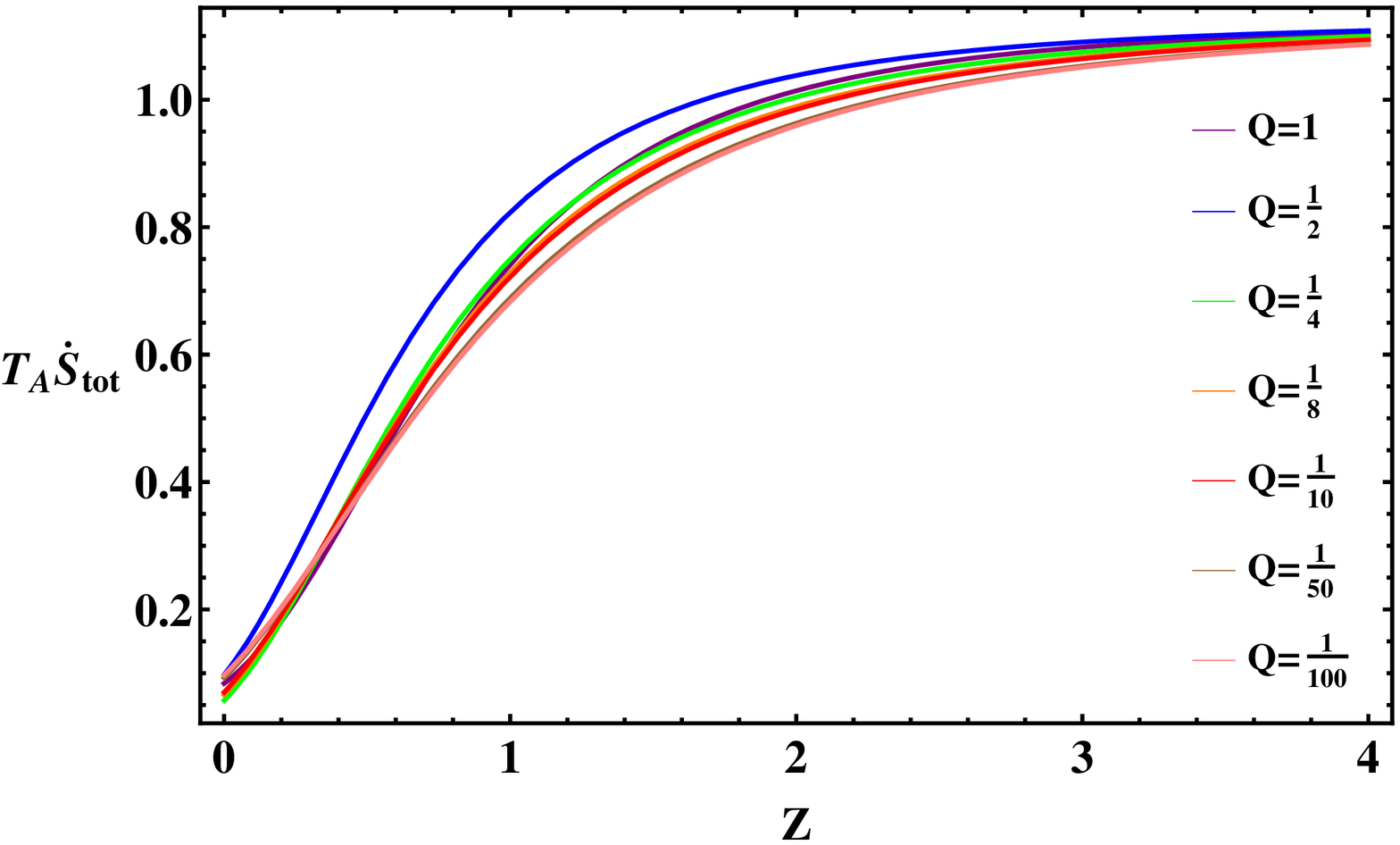}}\hspace{.1cm}
\subfigure[\label{fig-phi} ]{ \includegraphics[width=.48\textwidth]%
{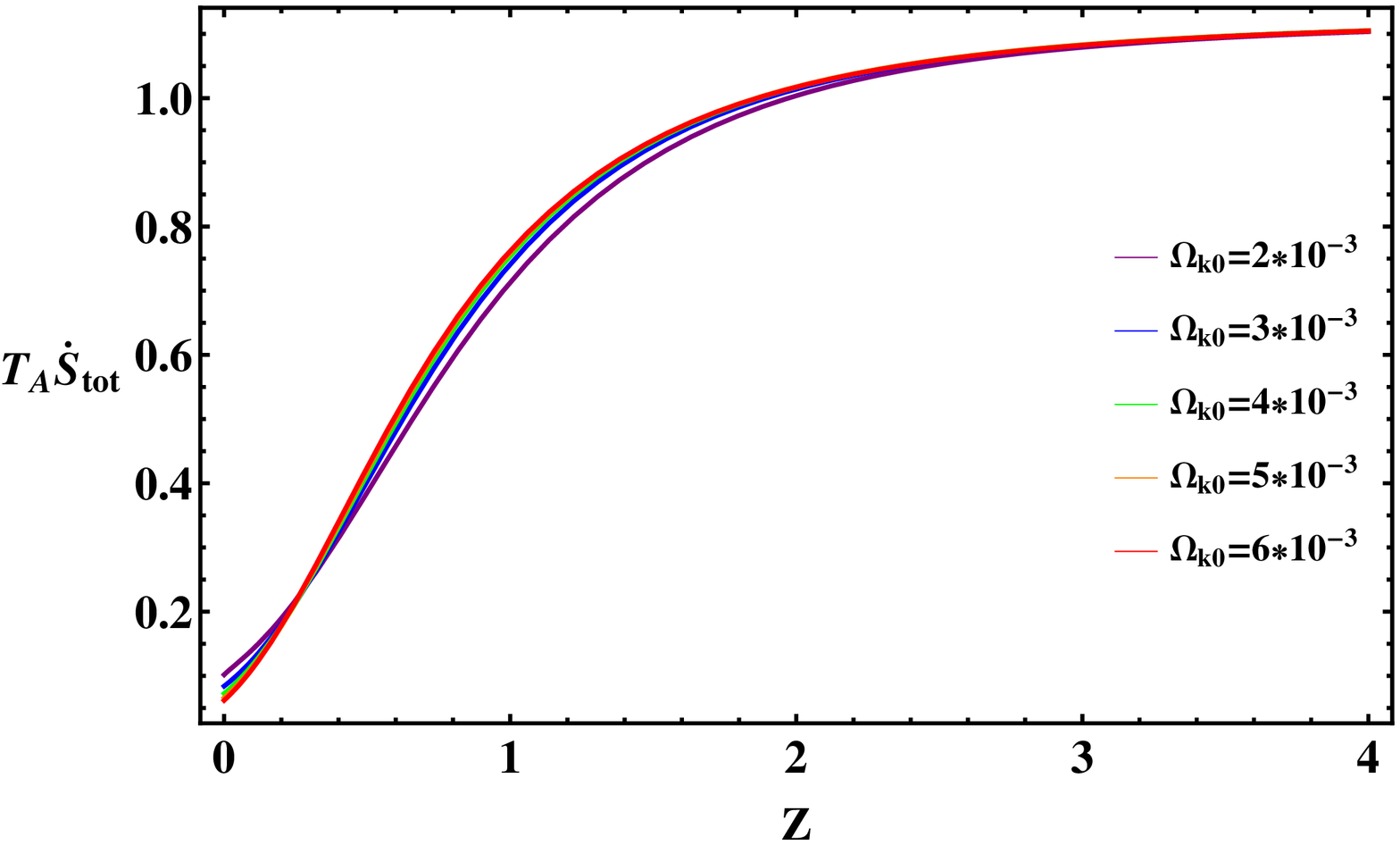}}
\end{minipage}
\caption{Evolutions of the matter entropy $T_A\dot{S}_m$, entropy of the apparent horizon $T_A\dot{S}_A$ and the GSL of thermodynamics $T_A\dot{S}_{tot}$ for GMG model with different values of $Q$ (with fixed $\Omega_{k_0}=3\times 10^{-3}$) and $\Omega_{k_0}$ (with fixed $Q=1$). Auxiliary parameters are $\Omega_{m_0}=0.3$, $\alpha_{3}=0$ and $\alpha_{4}=0.8$ \cite{Mic:2020}.
  }\label{fig3}
\end{figure}

Here, we are interested in examining the validity of GSL of thermodynamics for the GMG model. To do this, we first replace Eq. (\ref{L2}) into (\ref{s11_eq}) and  (\ref{s12_eq}) to obtain evolutions of the pressureless matter entropy $T_A\dot{S}_m$ (with $p_m=0$) and entropy of the apparent horizon $T_A\dot{S}_A$ as
\begin{equation}\label{s11_eq-GMG}
T_A\dot{S_m}=16\pi^2G H R_A^5 \rho_m\big[\rho_m-M_{P}^2 m^2 J (r-1)\xi\big]-4\pi H R_A^3\rho_m,
\end{equation}

\begin{equation}\label{s12_eq-GMG}
T_A\dot{S_{A}}=4 \pi H R_{A}^3\rho_m-8\pi^2 G H R_{A}^5\rho_m\big[\rho_m-M_{P}^2 m^2 J (r-1)\xi\big],
\end{equation}
where $\rho_m=\rho_{m_0}a^{-3}$. Finally, the GSL (\ref{st_eq}) in GMG theory takes the form
\begin{equation}\label{GSL-GMG}
T_{A} \dot{S}_{tot}= 8\pi^2 G H R_{A}^5\rho_m\big [\rho_m-M_{P}^2 m^2 J (r-1)\xi\big].
\end{equation}
In Fig. \ref{fig3}, using Eqs. (\ref{s11_eq-GMG}), (\ref{s12_eq-GMG}) and (\ref{GSL-GMG}) we plot evolutions of $T_A\dot{S}_m$, $T_A\dot{S}_A$ and $T_A\dot{S}_{tot}$ for different values of the GMG model parameters.
Figure \ref{fig3} shows that although the entropy of matter does not satisfy the second law of thermodynamics (i.e. $T_A\dot{S}_m<0$) in the near past, but when we add the entropy of horizon to the matter entropy, the GSL in GMG model is respected for different values of $Q$ and $\Omega_{k_0}$ during history of the universe.

According to \cite{Ken:2020}, it was pointed out that in the generalized massive gravity model due to having a positive cosmological
constant and stable perturbations, the region of
the parameter space $\alpha_4-\alpha_3$ (for the case $\alpha'_{2}>0$ in our model) should satisfy the following constraints
\begin{align}
&\alpha_3>-1,\nonumber\\
&\frac{1}{4}\left(3+2\alpha_3+3\alpha_3^2\right)<\alpha_4<1+\alpha_3+\alpha_3^2,\nonumber\\
&\alpha_3-\alpha_4+1>0.\label{PS}
\end{align}
The region where all conditions are satisfied has been plotted in Fig. 2 (left panel) of \cite{Ken:2020}. Following \cite{Setare,Basilakos}, we probed the validity of the GSL, $T_{A} \dot{S}_{tot}\geq 0$, for the allowed region (\ref{PS}) and found that the GSL is respected. In Fig. \ref{fig3}, we plot only the result for the case $\alpha_3=0$ and $\alpha_4=0.8$.

Regrading the $Q$ parameter, it should be noted that we only consider the values $Q\leqslant1$. This selection yields the cosmological solutions to be stable against linear perturbations \cite{Mic:2020}.

\section{dRGT massive gravity on de Sitter }\label{sec4}
In this section, the second model of generalized massive gravity namely dRGT massive gravity on de Sitter has been inspected. In this model, the secondary (fiducial) Minkowski metric in the standard dRGT is replaced by the de Sitter metric. The action of dRGT massive gravity on de Sitter is given by \cite{To:2011}
\begin{equation}\label{Lagrangian4}
S=M^2_P \int d^4 x \sqrt{-g}\left (\frac{R}{2}+m^2 U\right)+S_m,
\end{equation}
wherein $ m $ denotes the graviton mass and the dRGT potential terms have the following form
\begin{equation}\label{Lagrangian}
U=U_2+\alpha_3 U_3+\alpha_4 U_4,
\end{equation}
in which $U_2$, $U_3$ and $U_4$ are given by Eq. (\ref{un}). Also $\alpha_3$ and $\alpha_4$ are the free model parameters.

%

Following Langlois and Naruko \cite{Lan:2012}, we take the de Sitter metric $f_{ab}$ for the reference metric as follows
\begin{equation}\label{Lagrangian}
f_{ab} d \phi^a d \phi^b=-dT^2+b^2_{k}(T){\gamma_ {ij}}dX^i dX^j,
\end{equation}
where $\gamma_{ij}$ is the spatial metric and the de Sitter functions $b_{k}(T)$ (with $k=0,\pm1$) have the following form
\begin{equation}\label{Lagrangian}
b_{0}(T)=e^{H_cT}, \qquad b_{-1}(T)=H_c^{-1} \sinh(H_{c}T), \qquad b_{1}(T)=H_c^{-1} \cosh(H_{c}T).
\end{equation}
Note that for the case of $H_c\rightarrow 0 $, the Minkowski metric is recovered for the flat $b_0(T)=1$ and open $b_{-1}(T)=T$ universe and the latter case reduces to the Milne metric in the flat geometry.

The St\"{u}ckelberg fields $\phi^a$ are determined from the homogeneity and isotropy conditions as follows \cite{Lan:2012}
\begin{equation}
\phi^0=T=f(t), \qquad \phi^i=X^i=x^i.
\end{equation}
Therefore, the fiducial metric $f_{\mu\nu}$ corresponding to the de Sitter spacetime $f_{ab}$ yields
\begin{equation}\label{fmunudRGT}
f_{\mu\nu}\equiv f_{ab}\partial_{\mu}\phi^a\partial_{\nu}\phi^b={\rm Diag}\big[-\dot{f}^2,b_k^2\big(f(t)\big)\gamma_{ij}\big].
\end{equation}

Taking the variation of action (\ref{Lagrangian4}) with respect to the physical FRW metric $g_{\mu\nu}$, the modified Friedmann equations in the dRGT massive gravity on de Sitter can be obtained as
\begin{equation}\label{Lag1}
H^2+\frac {k}{a^2}=\frac {1}{3M_{P}^2} (\rho_m+\rho_{DE}),
\end{equation}
\begin{equation}\label{Lag2}
2\dot{H}+3H^2+\frac {k}{a^2}=-\frac {1}{M_{P}^2}(p_m+p_{DE}),
\end{equation}
where $\rho_{DE}$ and $p_{DE}$ are the effective energy density and pressure of dark energy associated with the massive graviton term defined as
\begin{align}\label{rhodRGT}
&\rho_{DE}=\frac{m^2 M_P^2}{a^3}(b_k(f)-a)\Big\lbrace(6+4\alpha_3+\alpha_4)a^2-(3+5\alpha_3+2\alpha_4)ab_k(f)\nonumber\\
&+(\alpha_3+\alpha_4)b^2_k(f)\Big\rbrace,
\end{align}
\begin{align}\label{pdRGT}
&p_{DE}=\frac{m^2 M_P^2}{a^2}\Big\lbrace\Big[6+4\alpha_3+\alpha_4-(3+3\alpha_3+\alpha_4)\dot{f}\Big]a^2\nonumber\\
&-2\Big[3+3\alpha_3+\alpha_4-(1+2\alpha_3+\alpha_4)\dot{f}\Big]ab_k(f)\nonumber\\
&+\Big[1+2\alpha_3+\alpha_4-(\alpha_3+\alpha_4)\dot{f}\Big]b^2_k(f)\Big\rbrace.
\end{align}
Varying the action (\ref{Lagrangian4}) with respect to the fiducial metric $f_{\mu\nu}$, Eq. (\ref{fmunudRGT}), one can get the equation of motion governing the St\"{u}ckelberg field $f(t)$ as follows
\begin{equation}\label{stuckdRGT}
\Big[(3+3\alpha_3+\alpha_4)a^2-2(1+2\alpha_3+\alpha_4)ab_k(f)+(\alpha_3+\alpha_4)b^2_k(f)\Big]\left(\dot{a}-\varepsilon_{f}\frac{d b_k(f)}{d f}\right)=0,
\end{equation}
where $\varepsilon_f$ denotes the sign of $f$. The St\"{u}ckelberg field equation (\ref{stuckdRGT}) has the three solutions, the first two ones $b_k\big(f(t)\big) =X_\pm a(t)$ represent the effective cosmological constant and are independent of the specific form of $ b_k(f)$. Here,
\begin{equation}
X_\pm=\frac{1+2\alpha_3+\alpha_4\pm\sqrt{1+\alpha_3+\alpha_3^2-\alpha_4}}{\alpha_3+\alpha_4}.
\end{equation}
The third solution of the St\"{u}ckelberg equation (\ref{stuckdRGT}) satisfy the following relation
\begin{equation}\label{ff}
\varepsilon_f\frac{d b_k(f)}{d f}=\dot{a}.
\end{equation}
For the case of flat FRW universe ($k=0$) which we focus on it in what follows, substituting the de Sitter function $b_0\big(f(t)\big)=e^{H_c f(t)}$ into Eq. (\ref{ff}) and assuming $\dot{f}>0$ (i.e. $\varepsilon_f=1$), the result yields the following St\"{u}ckelberg function
\begin{equation}\label{fdRGT}
f(t)=H_c^{-1}\ln\left(\frac{aH}{H_c}\right).
\end{equation}
Replacing the solution (\ref{fdRGT}) into Eqs. (\ref{rhodRGT}) and (\ref{pdRGT}) one can get
\begin{equation}\label{L3dRGT}
\rho_{DE}=-m^2\left(1-\frac{H}{H_c}\right)\Bigg[6+4\alpha_3+\alpha_4-(3+5\alpha_3+2\alpha_4)\frac{H}{H_c}+(\alpha_3+\alpha_4)\frac{H^2}{H_c^2}\Bigg],
\end{equation}
\begin{align}\label{L4dRGT}
&p_{DE}=m^2\Bigg[6+4\alpha_3+\alpha_4-(3+3\alpha_3+\alpha_4)\frac{H}{H_c}\left(3+\frac{\dot{H}}{H^2}\right)\nonumber\\
&+(1+2\alpha_3+\alpha_4)\frac{H^2}{H_c^2}\left(3+2\frac{\dot{H}}{H^2}\right)-(\alpha_3+\alpha_4)\frac{H^3}{H_c^3}\left(1+\frac{\dot{H}}{H^2}\right)\Bigg].
\end{align}
Adding Eqs. (\ref{L3dRGT}) and (\ref{L4dRGT}) one can get
\begin{align}\label{rhopdRGT}
&\rho_{DE}+p_{DE}= m^2\frac{\dot{H}}{H^2} \frac{H}{H_c} \Bigg[-(3+3\alpha_3+\alpha_4)+(2+4\alpha_3+2\alpha_4) \frac{H}{H_c}\nonumber\\
&-(\alpha_3+\alpha_4) \frac{H^2}{H_{c}^2}\Bigg].
\end{align}
Substituting Eqs. (\ref{L3dRGT}) and (\ref{L4dRGT}) into the Friedmann Eqs. (\ref{Lag1}) and (\ref{Lag2}) and using $\rho_m=\rho_{m_0} a^{-3}$ for the pressureless matter ($p_m=0$), in the case of flat universe ($k=0$) one can obtain
\begin{align}\label{HdRGT}
&\frac{m^2}{H_0^2}\frac{H}{H_c}\left[-\frac{1}{3}(\alpha_3+\alpha_4)\frac{H^2}{H_c^2}+(1+2\alpha_3+\alpha_4)
\frac{H}{H_c}-(3+3\alpha_3+\alpha_4)\right]\nonumber\\
&=-\frac{H^2}{H^2_0}+\Omega_{m_0}(1+z)^3-2\frac{m^2}{H_0^2}\left(1+\frac{2}{3}\alpha_3+\frac{1}{6}\alpha_4\right),
\end{align}
\begin{align}\label{HdotdRGT}
&\frac{\dot{H}}{H^2}\Bigg\lbrace-2\frac{H^2}{H_0^2}+\frac{m^2}{H_c^2}\frac{H}{H_0}
\Bigg[(3+3\alpha_3+\alpha_4)\frac{H_c}{H_0}-2(1+2\alpha_3+\alpha_4)\frac{H}{H_0}\nonumber\\
&+(\alpha_3+\alpha_4)\frac{H^2}{H_0H_c}\Bigg]\Bigg\rbrace=3\Omega_{m_0}(1+z)^3,
\end{align}
where $\Omega_{m_0}\equiv\rho_{m_0}/(3M_P^2H_0^2)$.

From Eqs. (\ref{L3dRGT}), (\ref{L4dRGT}), (\ref{HdRGT}) and (\ref{HdotdRGT}),  the effective equation of state parameter $\omega_{DE}$ for the dRGT massive gravity on de Sitter reads
\begin{equation}\label{omm_eq}
\omega_{DE}\equiv\frac{p_{DE}}{\rho_{DE}}=-1-\frac{2\Big(\frac{H}{H_0}\Big)^2\frac{\dot{H}}{H^2}
+3\Omega_{m_0}(1+z)^3}{3\Big[\Big(\frac{H}{H_0}\Big)^2-\Omega_{m_0}(1+z)^3\Big]}.
\end{equation}
Here, to obtain the dynamics of the Hubble parameter $H(z)$ we need to solve Eq. (\ref{HdRGT}). To do this, we concentrate on the special case including $\alpha_3=-\alpha_4$.

For the case $\alpha_3=-\alpha_4$, from Eq. (\ref{HdRGT}) one can get
\begin{equation}\label{e20_eq}
\frac{H(z)}{H_0}=\frac{-3\left(1+\frac{2\alpha_3}{3}\right)\frac{\beta_1}{\beta_2}+\sqrt{\Delta}}{2\Big[1-(1+\alpha_3)\frac{\beta_1}{\beta_2^2}\Big]},
\end{equation}
where
\begin{equation}\label{ee_eq1}
\Delta=9\left(1+\frac{2\alpha_3}{3}\right)^2\frac{\beta_1^2}{\beta_2^2}+4\Bigg[\beta_1(2+\alpha_3)+\Omega_{m_0}(1+z)^{3}\Bigg]\Bigg[1-\frac{\beta_1}{\beta_2^2}(1+\alpha_3)\Bigg],
\end{equation}
and $\beta_1$ and $\beta_2$ are the free model parameters defined as
\begin{equation}\label{beta12}
\beta_1\equiv-\frac{m^2}{H_0^2}, \qquad \beta_2\equiv\frac{H_c}{H_0}.
\end{equation}
Here, setting $H(z=0)=H_0$ into Eq. (\ref{e20_eq}) yields
\begin{equation}
\beta_1=\frac{\beta_2(1-\Omega_{m_0})}{(1+\alpha_3)\frac{1}{\beta_2}-(3+2\alpha_3)+2\beta_2(1+\frac{1}{2}\alpha_3)}.
\end{equation}
Here, we have only three free parameters including $\Omega_{m_0}$, $\alpha_3$ and $\beta_2$.
From Eq. (\ref{HdotdRGT}), the deceleration parameter $q(z)=-1-\dot{H}/H^2$ for the case of $\alpha_3=-\alpha_4$ can be obtained as follows
\begin{equation}\label{q2}
q(z)=-1+\frac{3\Omega_{m_0}(1+z)^3}{2\frac{H^2}{H_0^2}+\frac{\beta_1}{\beta_2^2}\frac{H}{H_0}\Big[3(1+\frac{2\alpha_3}{3})\beta_2-2\frac{H}{H_0}(\alpha_3+1)\Big]}.
\end{equation}

\begin{figure}[H]
\hspace*{-.5cm}
\begin{minipage}[b]{1\textwidth}
\subfigure[\label{fig-phi} ]{ \includegraphics[width=.51\textwidth]%
{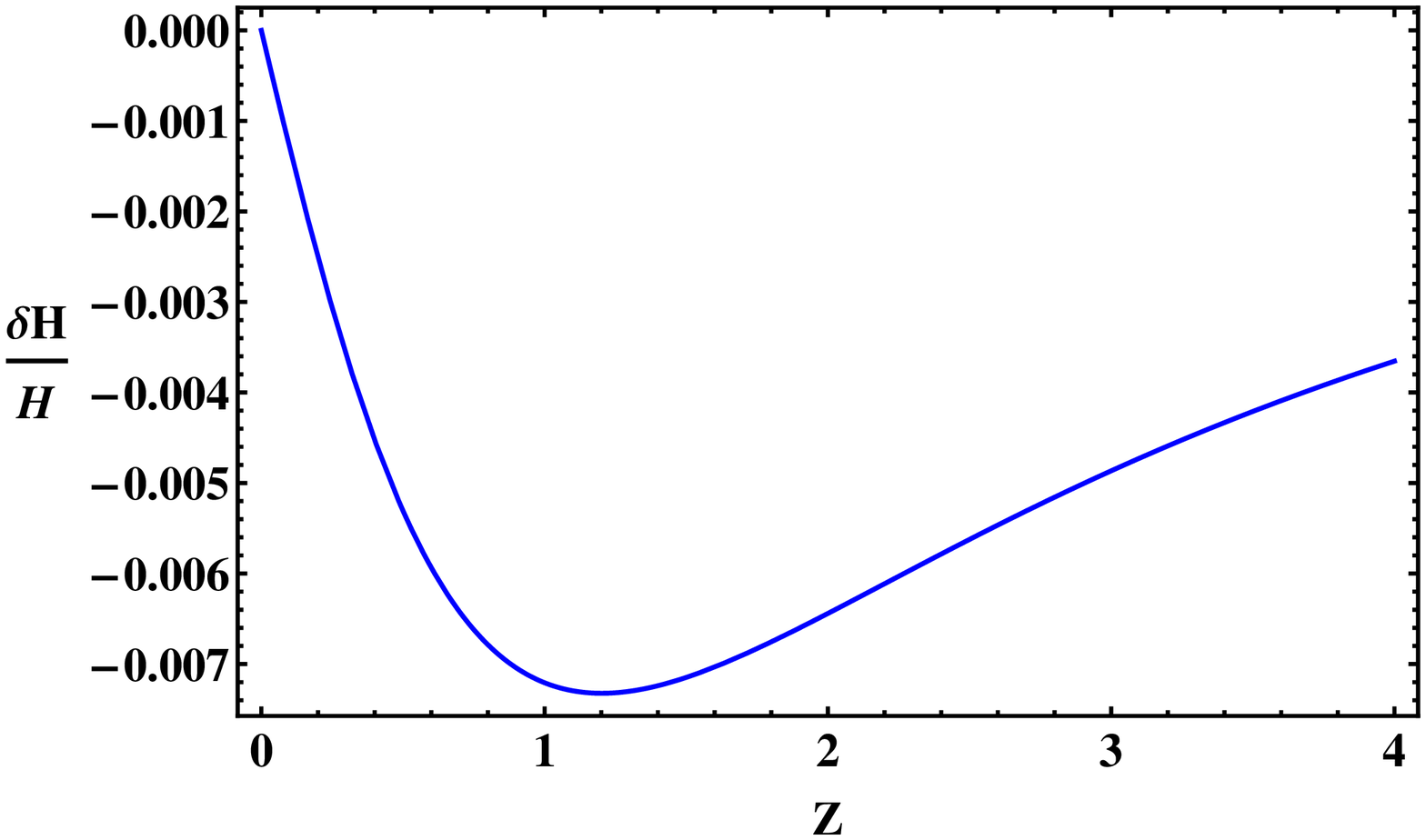}}\hspace{.05cm}
\subfigure[\label{fig-srp}]{ \includegraphics[width=.49\textwidth]%
{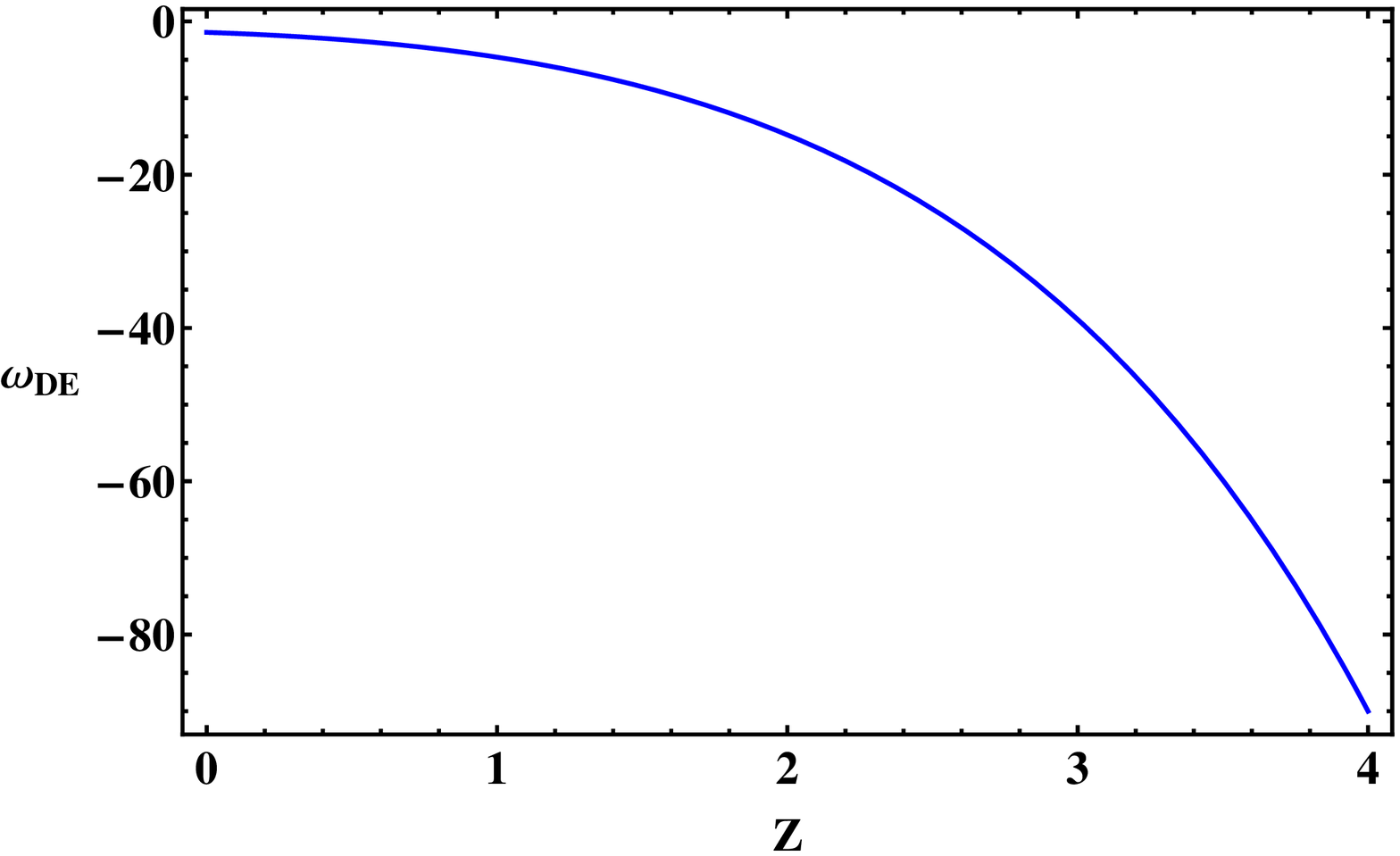}}\\
\subfigure[\label{fig-phi} ]{ \includegraphics[width=.50\textwidth]%
{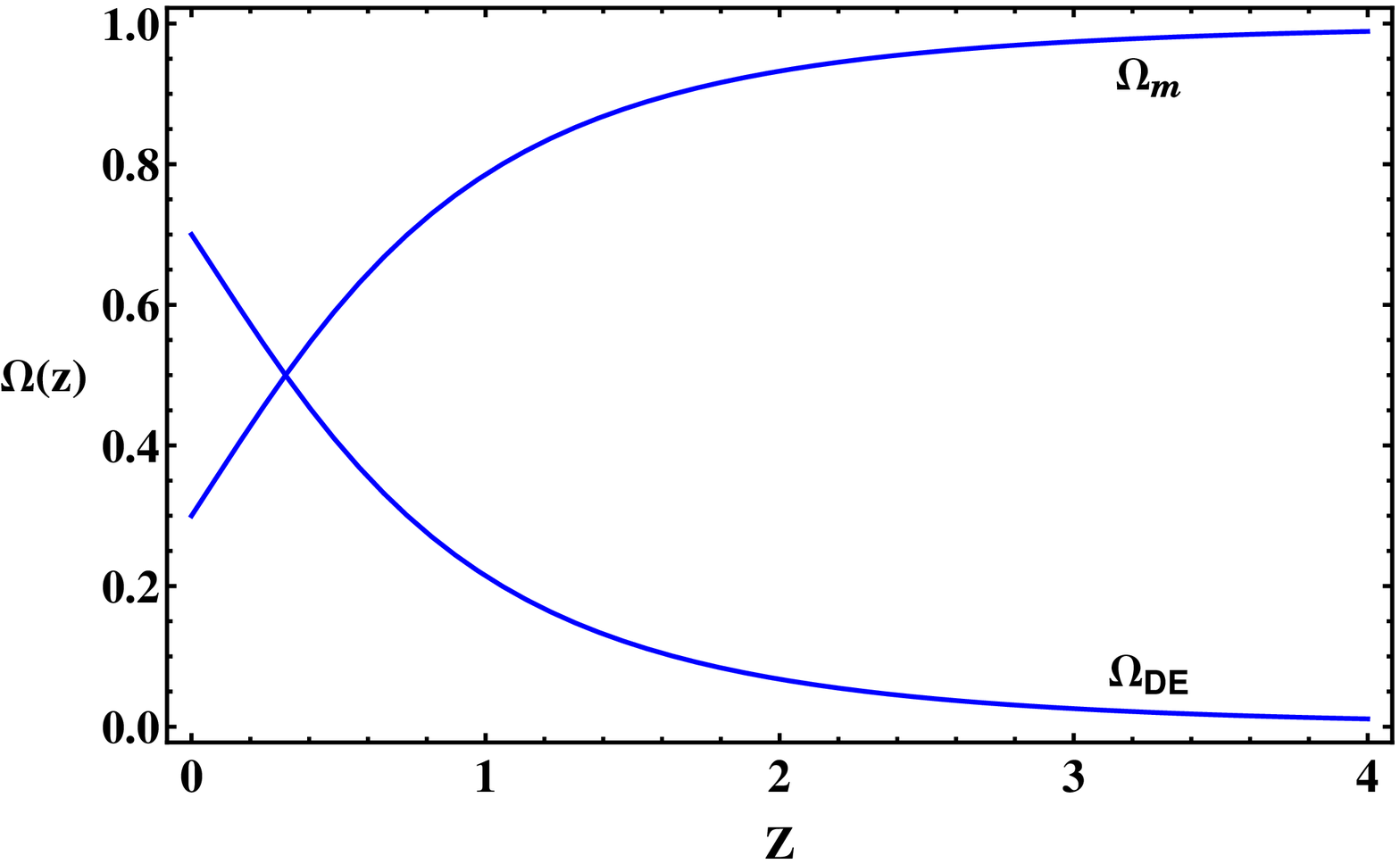}}\hspace{.05cm}
\subfigure[\label{fig-phi} ]{ \includegraphics[width=.49\textwidth]%
{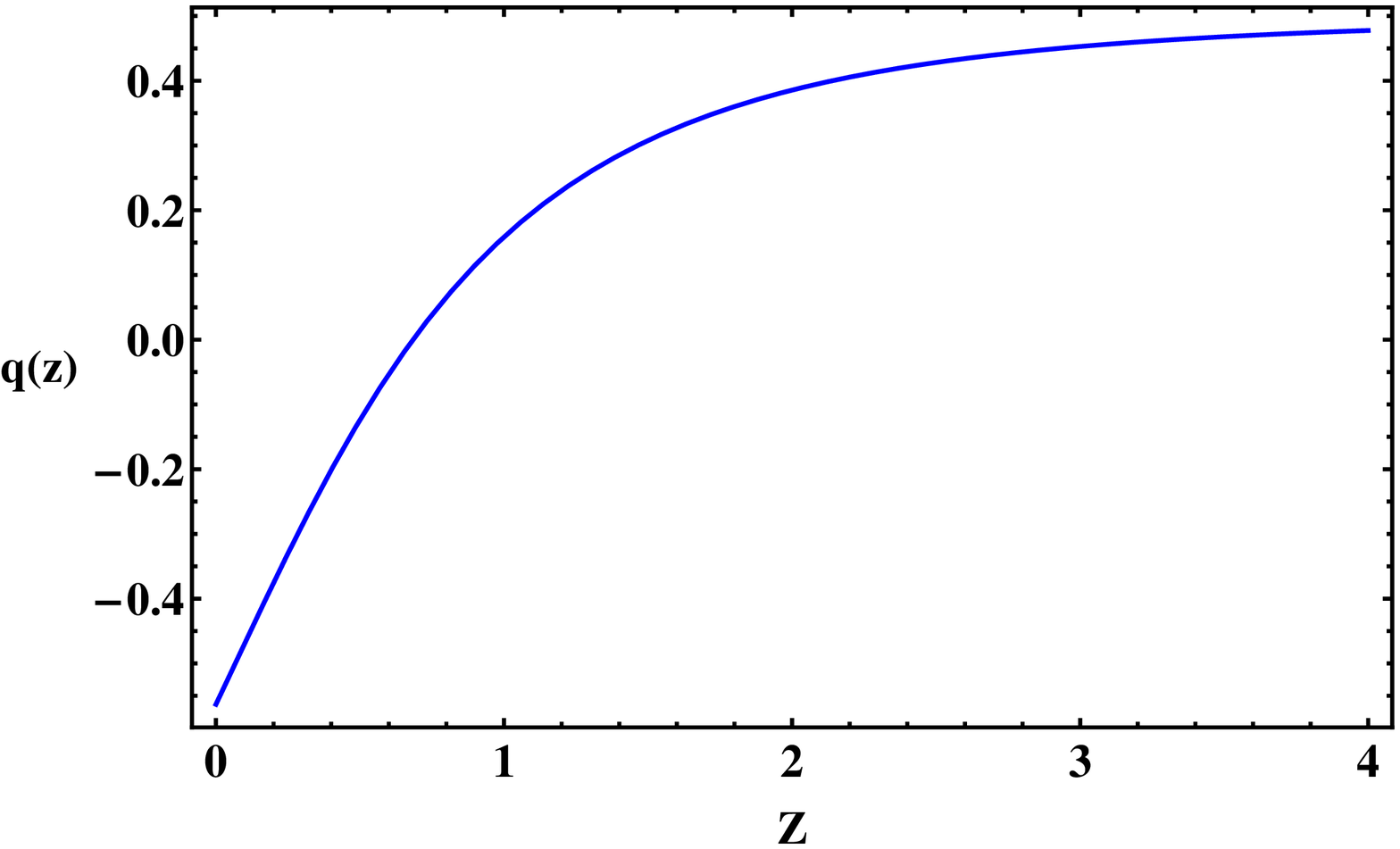}}
\end{minipage}
\caption{Variations of the fractional deviation of the Hubble parameter from $\Lambda$CDM, where $\frac{\delta H}{H}=\frac{H-H_{\Lambda \rm CDM}}{H_{\Lambda \rm CDM}}$, the equation of state parameter of dark energy $\omega_{DE}$, the density parameters  ($\Omega_{m}$,$\Omega_{DE}$) and the deceleration parameter $q(z)$ versus the redshift $z$ for the dRGT theory on de Sitter. Auxiliary parameters are $\Omega_{m_0}=0.3$, $\beta_2=20.1$ and $\alpha_3=0.95$ \cite{Gon:2013}. }\label{fig4}
\end{figure}

Using Eqs. (\ref{omm_eq}), (\ref{e20_eq}) and (\ref{q2}) we plot the evolutionary behaviors of $H(z)$, $\omega_{DE}(z)$, $\Omega_{m}(z)=\Omega_{m_0}(1+z)^3H_0^2/H^2$, $\Omega_{DE}(z)=1-\Omega_{m}(z)$ and $q(z)$ in Fig. \ref{fig4}.

Figure \ref{fig4} shows that: (i) fractional deviation of the Hubble parameter of dRGT model on de Sitter from $\Lambda$CDM is in order of $\mathcal{O}(10^{-3})$. (ii) The equation of state parameter of dark energy behaves like phantom dark energy (i.e. $\omega_{DE}<-1$) and tends to $\Lambda$CDM (i.e. $\omega_{DE}\rightarrow -1$) in the future. (iii) The density parameters $\Omega_m$ and $\Omega_{DE}$, respectively, start from 1 and 0 at early times and go toward their values at the present time. (iv) The deceleration parameter $q(z)$ begins from matter dominated era ($q=0.5$) and shows a transition from decelerating ($q>0$) to accelerating ($q<0$) phase in the near past.

\subsection{Generalized second law of thermodynamic in dRGT massive gravity on de Sitter}\label{subsec2}
Here, similar to what we did for the GMG model we try to check validity of the GSL of thermodynamics in the dRGT massive gravity on de Sitter.
To this aim, substituting Eq. (\ref{rhopdRGT}) into (\ref{s11_eq}) and  (\ref{s12_eq}) one can get

\begin{align}\label{s11_eq-dRGT}
T_A\dot{S_m}=&16\pi^2 G H R_A^5 \rho_m \Bigg\{\rho_m+m^2\frac{\dot{H}}{H^2} \frac{H}{H_c} \times  \Bigg[-(3+3\alpha_3+\alpha_4)+(2+4\alpha_3+2\alpha_4) \frac{H}{H_c}\nonumber\\
&-(\alpha_3+\alpha_4) \frac{H^2}{H_{c}^2}\Bigg]\Bigg\}  -4\pi H R_A^3\rho_m,
\end{align}

\begin{align}\label{s12_eq-dRGT}
T_A\dot{S_{A}}=& 4 \pi H R_{A}^3\rho_m-8\pi^2 G H R_{A}^5\rho_m\Bigg\{\rho_m+m^2 \frac{\dot{H}}{H^2} \frac{H}{H_c}\nonumber\\
&\times \Bigg[-(3+3\alpha_3+\alpha_4)
+(2+4\alpha_3+2\alpha_4) \frac{H}{H_c}
-(\alpha_3+\alpha_4) \frac{H^2}{H_{c}^2}\Bigg]\Bigg\},
\end{align}

where $\rho_m=\rho_{m_0}a^{-3}$. Finally, the GSL (\ref{st_eq}) in dRGT theory on de Sitter takes the form

\begin{align}\label{st_eq-dRGT}
T_{A} \dot{S}_{\rm tot}=& 8\pi^2 G H R_{A}^5 \rho_m\Bigg\{\rho_m+m^2\frac{\dot{H}}{H^2} \frac{H}{H_c} \times  \Bigg[-(3+3\alpha_3+\alpha_4)+(2+4\alpha_3+2\alpha_4) \frac{H}{H_c}\nonumber\\
&-(\alpha_3+\alpha_4) \frac{H^2}{H_{c}^2}\Bigg]\Bigg\}.
\end{align}

\begin{figure}[H]
\begin{minipage}[b]{1\textwidth}
\subfigure[\label{fig-phi} ]{ \includegraphics[width=.48\textwidth]%
{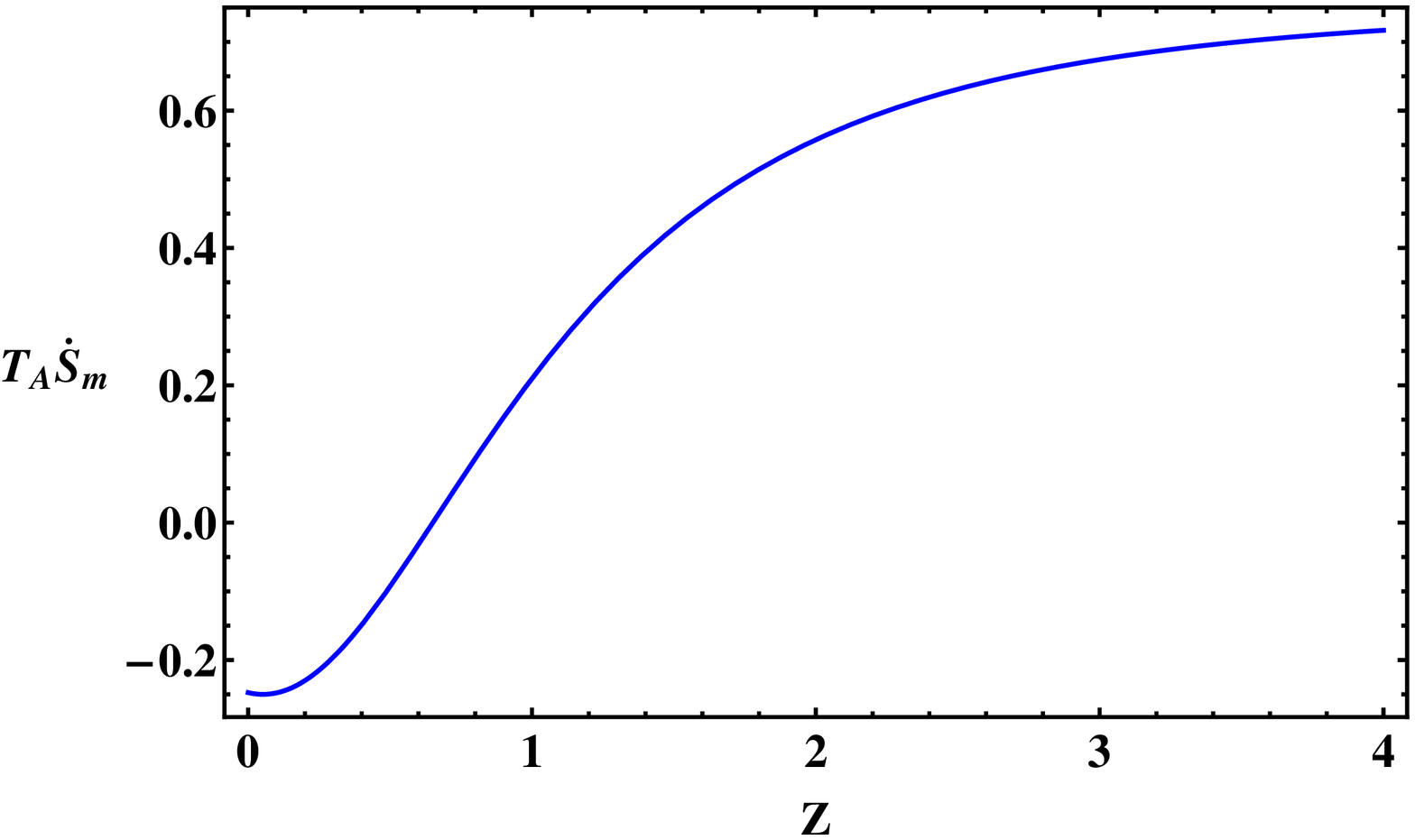}}\hspace{.1cm}
\subfigure[\label{fig-srp}]{ \includegraphics[width=.48\textwidth]%
{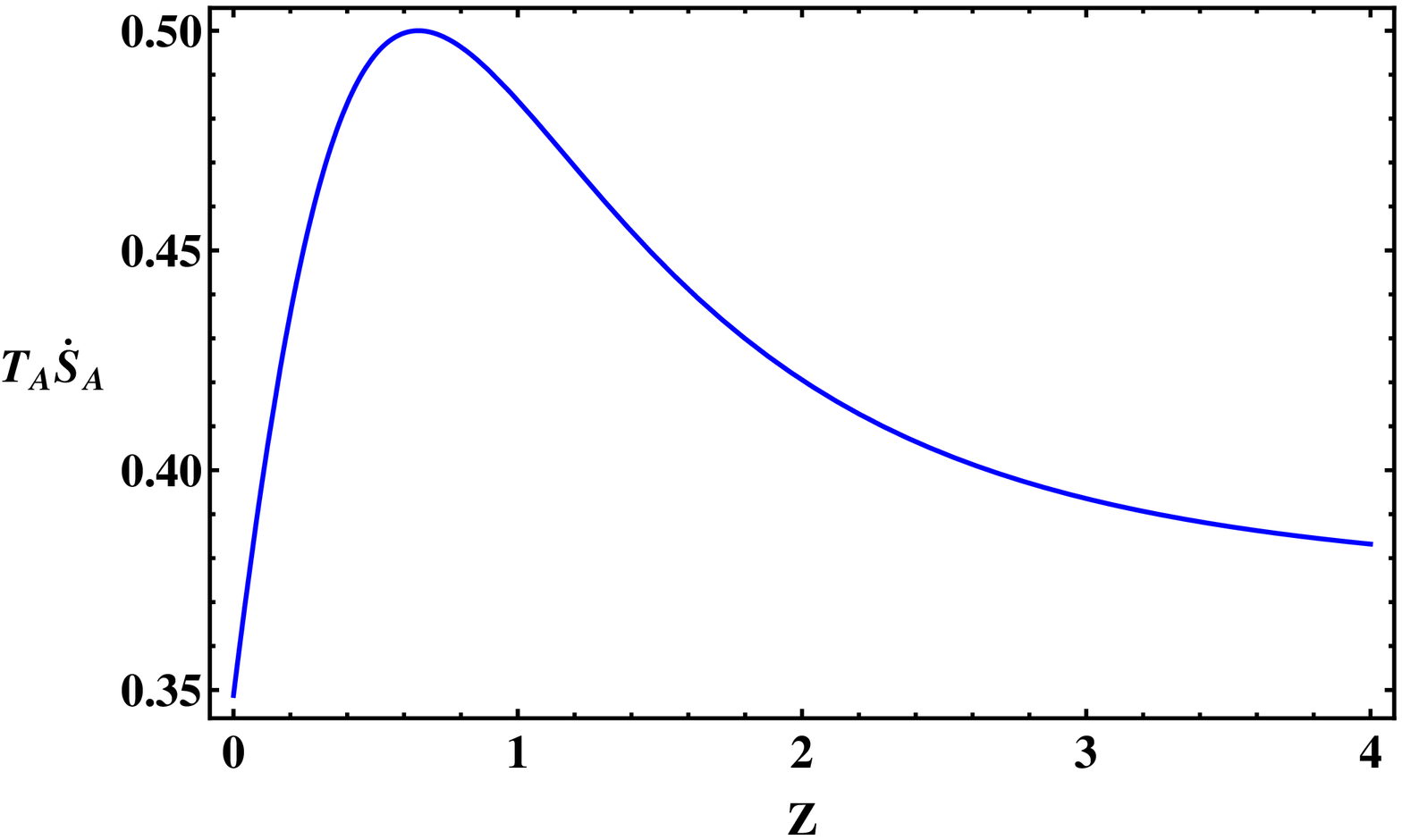}}\\
\subfigure[\label{fig-phi} ]{ \includegraphics[width=.49\textwidth]%
{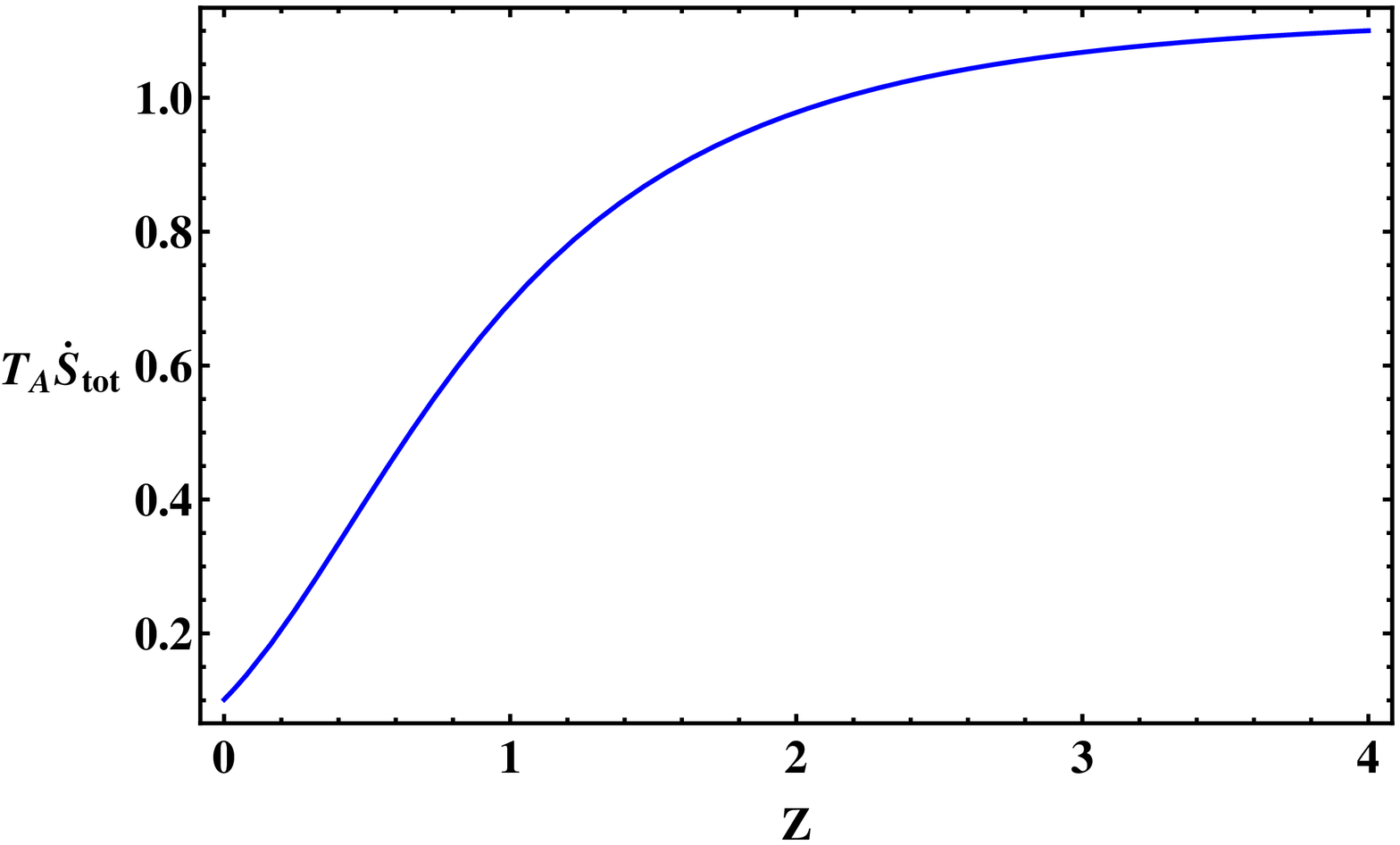}}\centering
\end{minipage}
\caption{Variations of the matter entropy $T_A\dot{S}_m$, entropy of the apparent horizon $T_A\dot{S}_A$ and the GSL of thermodynamics $T_A\dot{S}_{tot}$ versus the redshift $z$ for the dRGT massive gravity on de Sitter. Auxiliary parameters are $\Omega_{m_0}=0.3$, $\beta_2=20.1$ and $\alpha_3=0.95$ \cite{Gon:2013}.}\label{fig6}
\end{figure}

Using Eqs. (\ref{s11_eq-dRGT}), (\ref{s12_eq-dRGT}) and (\ref{st_eq-dRGT}) we plot the evolution of matter, horizon and total entropy in dRGT model on de Sitter for the case of $\alpha_3=-\alpha_4$ in Fig. \ref{fig6}. This figure shows that the entropy of matter violates the second law of thermodynamics (i.e. $T_A\dot{S}_m<0$) for the region of $z<1$. But when we add the horizon entropy to the matter entropy, the total entropy of the universe in dRGT massive gravity on de Sitter satisfies the GSL of thermodynamics.

\section{conclusions}\label{sec5}
Here, we investigated the GSL of thermodynamics within the framework of massive gravity. According to the GSL, the time evolution of matter entropy and horizon entropy must be increasing function of time. To this aim, we considered a FRW universe filled with the pressureless matter and enclosed by the apparent horizon. In addition, we obtained a generalized formula for the GSL which is applicable for modified gravity theories. In the next, we considered two massive gravity models including the generalized massive gravity and dRGT theory on de Sitter. The GMG model is a generalized version of the standard dRGT in which the mass graviton is slow varying time dependent. In dRGT on de Sitter, the secondary (fiducial) Minkowski metric in standard dRGT is replaced by the de Sitter reference metric. In the next, we first studied the cosmological background for the GMG and dRGT on de Sitter models and then we examined the GSL for both of them with different model parameters. Our results show that:

\begin{itemize}
\item For the GMG model, $\delta H/H$ shows deviation from the $\Lambda$CDM for different model parameters $Q$ and $\Omega_{k_0}$. Also for the case of $Q\rightarrow 0$, the result of standard dRGT model is recovered (i.e.  $\xi\rightarrow \xi_{\rm dRGT}=2.80902$).
\item For the GMG and dRGT on de Sitter models, the fractional deviation of the Hubble parameter $\delta H/H$ from $\Lambda$CDM, respectively are in order of $\mathcal{O}(10^{-2})$ and $\mathcal{O}(10^{-3})$.
\item For the both GMG and dRGT on de Sitter:  (i) the equation of state parameter behaves like phantom dark energy (i.e. $\omega_{DE}<-1$). (ii) The density parameters $\Omega_m$ and $\Omega_{DE}$ start from 1 and 0 at early time and reach to their values at the present. (iii) The deceleration parameter begins from matter dominated epoch ($q=0.5$) and shows a transition from decelerating ($q>0$) to accelerating ($q<0$) phase near the past. (iv) The entropy of matter violates the second law of thermodynamics (i.e. $T_A\dot{S}_m<0$) but when we add the horizon entropy to the matter entropy, the total entropy satisfies the GSL for both models.
\end{itemize}
\subsection*{Acknowledgements}
The authors thank the referee for his/her valuable comments.


\end{document}